\documentclass{article}

\usepackage[letterpaper,top=2cm,bottom=2cm,left=3cm,right=3cm,marginparwidth=1.75cm]{geometry}

\usepackage{amsmath}
\usepackage{amsfonts}
\usepackage{amssymb}
\usepackage{bm}
\usepackage{graphicx}
\usepackage{float}
\usepackage{subcaption}
\usepackage{lipsum} 
\usepackage[colorlinks=true, allcolors=blue]{hyperref}

\title{\Large \textsc{Combining Hyperbolic Quadrature Method of Moments and
Discrete-Velocity-Direction Models for Solving BGK-type Equations}}
\author{\large TIANSHU LI$^{1, \dagger}$, YIHONG CHEN$^{2, \dagger}$, AND QIAN HUANG$^{3, *}$}
\date{}

\begin{document}
\maketitle
\renewcommand{\thefootnote}{}
\footnotetext{$^1$ Weiyang College, Tsinghua University, Beijing, 100084, China.\quad \textit{Email address:} \href{mailto:lts21@mails.tsinghua.edu.cn}{lts21@mails.tsinghua.edu.cn}
}

\footnotetext{$^2$ Department of Mathematical Sciences, Tsinghua University, Beijing, 100084, China.\quad \textit{Email address:} \href{mailto:chenyiho20@mails.tsinghua.edu.cn}{chenyiho20@mails.tsinghua.edu.cn}
}
\footnotetext{$^3$ Institute of Applied Analysis and Numerical Simulation, University of Stuttgart, Stuttgart, 70569, Germany.\quad \textit{Email address:} \href{mailto:qian.huang@mathematik.uni-stuttgart.de}{qian.huang@mathematik.uni-stuttgart.de}
}
\footnotetext{$^{\dagger}$ These authors contributed equally to this work.}
\footnotetext{$^{*}$ Corresponding author.}
\begin{abstract}
This paper introduces the discrete-velocity-direction model (DVDM) in conjunction with the hyperbolic quadrature method of moments (HyQMOM) to develop a multidimensional spatial-temporal approximation of the BGK equation, termed DVD-HyQMOM. Serving as a multidimensional extension of HyQMOM, DVD-HyQMOM model achieves higher accuracy than other DVDM submodels, especially with an increased number of abscissas. The efficiency and effectiveness of this model are demonstrated through various numerical tests.
\end{abstract}

\section{Introduction}

Kinetic theories and the Boltzmann equation form the foundation for studying non-equilibrium dynamics in many-body systems, traditionally in rarefied gas flows but also in emerging areas such as aerosol behavior \cite{yao17}, multiphase flows \cite{Friedlander00}, and active matter systems \cite{Virgillito21}. These fields involve complex kinetic processes that describe particle interactions and transport, with applications ranging from vacuum technologies \cite{Sharipov07} to ultrafine particle dynamics \cite{Phillips75} and lunar-lander plume analysis \cite{Trafton11}. However, solving the Boltzmann equation directly is computationally challenging due to the complexity of collision terms and the high-dimensional nature of distribution functions involving space, velocity, and other properties.

To address these difficulties, simplified models have been developed. Among them, the Bhatnagar–Gross–Krook (BGK) model approximates the collision term as a relaxation process toward local equilibrium, retaining essential conservation laws and the hydrodynamic limit \cite{krook54, Pareschi14, Harris14, Perthame89}. Yet, handling the nonlinearity and non-local dependence of the Maxwellian distribution remains an obstacle, especially as computational cost rises significantly in high dimensions, especially if we discretize the phase space directly. 

In order to tackle the high dimension computational cost issue, the method of moments, especially the quadrature-based methods of moments (QBMM) techniques \cite{fox2008, fox2017} born out of necessity, reducing velocity dependence and generating hydrodynamic equations with macroscopic variables. Among them, the quadrature method (QMOM) relies on nonlinear reconstructions of distributions that can be far from equilibrium and preserves positivity. In theory, this approach is valid for all realizable moments. The centers are determined as the roots of an orthogonal polynomial derived from these moments. This makes the QMOM computationally efficient, leading to its popularity in particulate flow simulations \cite{fox13, qian2020}. However, the system generated by QMOM is non-hyperbolic, which can result in unphysical shock formations in simulations \cite{Massot12, fox2008, yong2020}. Then, the Gaussian-extended quadrature method of moments (Gaussian-EQMOM), reconstructing the velocity distribution as a weighted sum of Gaussian functions, stands out and the hyperbolicity is proven in \cite{fox2017, yong2020}. However, it is difficult to solve the nodes by enlarging the order of the Gaussian-EQMOM, which greatly reduces the accuracy and applicability. This led to the development of hyperbolic quadrature method of moments (HyQMOM) \cite{fox22, Laboureur21}, which preserves hyperbolicity and, as a sum of Dirac delta functions, extends well to higher number of abscissas, avoiding all the aforementioned drawbacks.

Aiming at a higher-dimensional version of HyQMOM, we introduced a novel approach, discrete-velocity-direction model (DVDM) \cite{chen1, chen2}, by allowing particle transport in finite, fixed directions while retaining continuous speed. This semi-continuous model provides an innovative framework for selecting discrete velocity nodes, distinct from the standard discrete-velocity methods that typically use uniform lattices. 

In this paper, we combine the Discrete-Velocity-Direction Model (DVDM) with collisions of BGK-type proposed in our previous projects \cite{chen1, chen2} and 1-D hyperbolic quadrature method of moments (HyQMOM), presenting a robust model, DVD-HyQMOM. It is a novel multidimensional extension of HyQMOM, preserving the advantages of 1-D HyQMOM over other moment methods. The performance of our DVD-HyQMOM model is examined with a series of 1-D and 2-D problems covering a wide range of flow regimes, including 1-D and 2-D Riemann problems and Couette flow problems. The comparison between this model and other DVDM submodels, such as DVD-EQMOM \cite{chen2}, as well as its performance with increasing abscissas selection, are also presented in the article. The numerical results present that our DVD-HyQMOM model is a well-developed model suitable for addressing high-dimensional problems. 

The rest of the paper is organized as follows. We provide a basic introduction to the BGK equation, equilibrium state, and macroscopic quantities in Section \ref{sec: bgk}. Section \ref{sec: dvdm} introduces the discrete-velocity-direction Models (DVDM). Section \ref{sec: stmodel} develops the DVD-HyQMOM model. The Neumann boundary condition and solid wall condition satisfying diffuse-scattering law are given in Section \ref{sec: bound}. In Section \ref{sec: alg}, we present the numerical schemes for DVD-HyQMOM, together with the methods for calculating the abscissas and weights. Numerical results are shown in Section \ref{sec: ex}. Finally, some conclusions are presented in Section \ref{sec: con}.

\section{BGK equation} \label{sec: bgk}
The BGK equation \cite{krook54} for the density function $f = f(t, \bm{x}, \bm{\xi}, \bm{\zeta})$ is

\begin{equation}
	\partial_{t}f + \bm{\xi}\cdot \nabla_{\bm{x}}f = \dfrac{1}{\tau}(\mathcal{E}[f] - f).
\end{equation}

Let $\bm{x}\in \mathbb{R}^{D} (D = 2, 3)$ be the spatial position and $\bm{\xi}\in \mathbb{R}^{D}$ be the molecule velocity. It is important to take $\bm{\zeta \in \mathbb{R}^{L}}$ into consideration, which represents the possible internal molecular degrees of freedom \cite{xu15}, and $\tau$ is a characteristic collision time.

The local equilibrium state $\mathcal{E}[f] = \mathcal{E}[f](t, \bm{x}, \bm{\xi}, \bm{\zeta})$ is the following Maxwell distribution

\begin{equation}
	\mathcal{E}[f] = \dfrac{\rho}{(2\pi \theta)^{(D+L)/2}}\exp\left(-\dfrac{|\bm{\xi} - \bm{U}|^2 + |\bm{\zeta}|^2}{2\theta}\right)
\end{equation}

Here the $\left|\cdot\right|$ presents the Euclidean length of a vector. The classical macroscopic fluid quantities we mostly focus on include density $\rho$, velucity $\bm{U}$, temperature $\theta$ and pressure $p$, and they are defined by

\begin{align}
\rho &= \left< f\right>,\quad  \bm{U} = \dfrac{\left< \bm{\xi}f\right>}{\rho} \in \mathbb{R}^{D},\notag\\
 E &= \dfrac{\left| \bm{U}\right|^2  + (D+L)\theta }{2} = \dfrac{1}{\rho}\left< \dfrac{\left| \bm{\xi}\right|^2  + \left| \bm{\zeta}\right|^2 }{2}f\right>,\quad p = \rho \theta
\end{align}

where the bracket $\left<\cdot\right>$ is defined as the integral $\left<g\right> = \int_{\mathbb{R}^{L}}\int_{\mathbb{R}^{D}} g(\bm{\xi}, \bm{\zeta}) \text{d}\bm{\xi} \text{d}\bm{\zeta}$. 

 Here the equilibrium distribution $\mathcal{E}[f]$ satisfies the following equations

  \begin{align}
  \left\{ 
  \begin{array}{ll}
    \hfill \mathcal{E}[f] &= \exp(\bm{\alpha}_{eq} \cdot \bm{m}(\bm{\xi}, \bm{\zeta}))\\
    \hfill \bm{\alpha}_{eq} &= \left(\ln \dfrac{\rho}{(2\pi \theta)^{(D + L)/2}}-\dfrac{\left|\bm{U}\right|^2}{2\theta}, \dfrac{\bm{U}}{\theta}, -\dfrac{1}{\theta}\right)^{T}\\
   \hfill \bm{m}(\bm{\xi}, \bm{\zeta}) &= \left(1, \bm{\xi}, \dfrac{\left|\xi \right|^{2} + \left|\zeta \right|^{2}}{2}\right)^{T}
    \end{array}
    \right.
  \end{align}

According to the basic laws of conservation of those macroscopic quantities, $\mathcal{E}[f]$ should satisfy 
\begin{equation}
    \left<\mathcal{E}[f] \bm{m}(\bm{\xi}, \bm{\zeta})\right> = (\rho, \rho \bm{U}, \rho E)^{T} := \bm{\rho}
\end{equation}

 Another significant property $\mathcal{E}[f]$ should conform to is that $\mathcal{E}[f]$ is the unique non-negative minimization of the following kinetic entropy
\begin{equation}
    H\left[f\right] = \left<f\ln f  - f\right>
\end{equation}

 given any $\bm{\rho}$ with $\rho > 0$ and $\theta > 0$.


\section{Discrete-Velocity-Direction Models (DVDM)} \label{sec: dvdm}
As discussed in the previous work \cite{chen1, chen2}, the core idea of the Discrete-Velocity-Direction Model (DVDM) is to discretize and limit the directions of molecule transport, with the selected directions $\{\bm{l}_{m}\}$ uniformly distributed in space. For example, when $D = 2$, this model can be realized by setting $\bm{l}_{m} = (\cos\gamma_{m}, \sin\gamma_{m})$, with $\gamma_{m} = \frac{(m-1)\pi}{N} \text{or} \frac{(2m-1)\pi}{2N}$. It should be noted that along each direction, the velocity $\xi$ is continuously distributed, and the distribution function should take the weight function $\left|\xi\right|^{D-1}$ into consideration.

After selecting the $N$ directions, then the distribution  $f = f(t, \bm{x}, \bm{\xi}, \bm{\zeta})$ can be replaced by $N$ distributions \\$\left\{f_{m}(t, \bm{x}, \xi, \bm{\zeta})\right\}_{m=1}^{N}$, where $\xi \in \mathbb{R}, \bm{\zeta} \in \mathbb{R}^{L}$. Then the molecule velocity for $f_m = \left\{f_{m}(t, \bm{x}, \xi, \bm{\zeta})\right\}$ is $\xi \bm{l}_{m}$ and the BGK equation for $f_m$ becomes 
\begin{equation}
	\partial_{t} f_m + \xi \bm{l}_m\cdot \nabla_{\bm{x}} f_m =  \dfrac{1}{\tau}(\mathcal{E}_m - f_m), \quad m = 1, 2, ..., N
 \label{eq: DVD-BGK}
\end{equation}
with the local equilibriums $\mathcal{E}_{m} = \mathcal{E}_{m}(t, \bm{x}, \xi, \bm{\zeta})$.

Then we use the $N$ distributions $\left\{f_{m}(t, \bm{x}, \xi, \bm{\zeta})\right\}_{m=1}^{N}$ to define new fluid quantities, never forget the weight function $\left| \xi\right|^{D-1}$:
\begin{align}
	\rho &= s\sum_{m=1}^{N}\int_{\mathbb{R}}\int_{\mathbb{R}^{L}}f_{m}\left|\xi\right|^{D-1}\text{d}\bm{\zeta}\text{d}\xi\\  \rho \bm{U} &= s\sum_{m=1}^{N}\int_{\mathbb{R}}\int_{\mathbb{R}^{L}}\xi \bm{l}_{m}f_{m}\left|\xi\right|^{D-1}\text{d}\bm{\zeta}\text{d}\xi\notag\\
	\rho E &= s\sum_{m=1}^{N}\int_{\mathbb{R}}\int_{\mathbb{R}^{L}}\dfrac{\xi^2 + \left|\bm{\zeta}\right|^2}{2}f_{m}\left|\xi\right|^{D-1}\text{d}\bm{\zeta}\text{d}\xi.
\end{align}

To deal with the equilibrium term on the RHS of equation \ref{eq: DVD-BGK}, we apply the conservation law, which means 
\begin{equation}
 \sum_{m=1}^{N}\int_{\mathbb{R}}\int_{\mathbb{R}^{L}}(1, \xi\bm{l}_{m}, \dfrac{\xi^2 + \left|\bm{\zeta}\right|^2}{2})\mathcal{E}_{m}\left|\xi\right|^{D-1}\text{d}\bm{\zeta}\text{d}\xi = (\rho, \rho\bm{U}, \rho E).	
 \label{eq: in+tr}
\end{equation}

The $\mathcal{E}_{m}$ has the variable-separating form $\mathcal{E}_{m}(t, \bm{x}, \xi, \bm{\zeta}) = \mathcal{E}_{tr, m}(t, \bm{x}, \xi)\mathcal{E}_{in, m}(t, \bm{x}, \bm{\zeta})$, in which the internal part $\mathcal{E}_{in, m}(t, \bm{x}, \bm{\zeta})$ is

\begin{equation}
	\mathcal{E}_{in, m}(t, \bm{x}, \bm{\zeta}) = \dfrac{1}{\sqrt{(2\pi \theta )^{L}}}\exp\left(-\dfrac{\left|\bm{\zeta}\right|^{2}}{2\theta}\right).
\end{equation}

Here the equilibrium temperature $\theta$ is

\begin{equation}
    \theta = \dfrac{2E - \left|\bm{U}\right|^{2}}{D + L}.
    \label{eq: temp}
\end{equation}

Put the variable-separating form of the local equilibrium into equation \Ref{eq: in+tr}, we derive constraints for the transport part $\mathcal{E}_{tr, m}(t, \bm{x}, \xi)$:

\begin{equation}
	\sum_{m=1}^{N}\int_{\mathbb{R}}\int_{\mathbb{R}^{L}}\bm{m}_{m}\mathcal{E}_{tr, m}\left|\xi\right|^{D-1}\text{d}\bm{\zeta}\text{d}\xi = \left(\rho, \rho\bm{U}, \rho (E-\dfrac{L}{2}\theta)\right) =: \bm{\rho}_{tr}, 	
\end{equation}

in which $\bm{m}_{m}(\xi) = (1, \xi\bm{l}_{m}, \xi^{2}/2)\in \mathbb{R}^{D+2}$. 

 For the transport part, we use the theorem proven in our previous paper \cite{chen1}.\\
\textbf{Theorem} \textit{Given} $\bm{\rho}_{tr} \in \mathbb{R}^{D+2}$ \textit{satisfying} $0<\left|\bm{\rho}_{tr}\right| < \infty$\textit{, if there exists} $\{g_m(\xi)\ge 0\}^N_{m=1}$ \textit{such that} 
\begin{equation*}
    s\sum_{m=1}^{N} \int_{\mathbb{R}} \bm{m}_m g_m \left|\xi\right|^{D-1} \text{d}\xi = \bm{\rho}_{tr}, 
\end{equation*}
\textit{then the discrete kinetic entropy }
\begin{equation*}
    H\left[\{g_m\}_{m=1}^{N}\right]:= s\sum_{m=1}^{N}\int_{\mathbb{R}}\int_{\mathbb{R}^{L}} (g_m \ln g_m - g_m)\left|\xi\right|^{D-1} \text{d}\bm{\zeta}\text{d}\xi
\end{equation*}
\textit{has a unique minimizer} $\{\mathcal{E}_{tr, m}\}_{m=1}^{N}$\textit{, and the minimizer has the exponential form }
\begin{equation*}
    \mathcal{E}_{tr, m} = \exp(\bm{\alpha}\cdot \bm{m}_{m}).
\end{equation*}
\textit{Here} $\bm{\alpha} = (\alpha_0, \bm{\hat{\alpha}}, \alpha_{D+1}) \in \mathbb{R}^{D+1} \times \mathbb{R}^{-}$ \textit{is the unique minimizer of the convex function}
\begin{equation*}
    J(\bm{\alpha}) := s\sum_{m=1}^{N} \int_{\mathbb{R}}\exp (\bm{\alpha} \cdot \bm{m}_m)\left|\xi\right|^{D-1} \text{d}\xi - \bm{\rho}_{tr}\cdot \bm{\alpha}.
\end{equation*}

 According to this theorem, we know $\mathcal{E}_{tr, m}$ is in the form of Gaussian distribution:
 
\begin{equation}
	\mathcal{E}_{tr, m} = \exp(\bm{\alpha}\cdot\bm{m}_{m}) = \dfrac{\rho_{m}}{\sqrt{2\pi \sigma^2}}\exp\left(-\dfrac{(\xi - u_{m}^2)}{2\sigma^2}\right),
 \label{eq: local eq}
\end{equation}

and the parameters $\rho_m, u_m, \sigma$ are related to $\bm{\alpha} = (\alpha_0, \bm{\hat{\alpha}}, \alpha_{D+1})$ as:

\begin{equation}
    \sigma^2 = -\dfrac{1}{\alpha_{D+1}}, u_m = \left(\bm{\hat{\alpha}} \cdot \bm{l}_m\right)\sigma^2, \rho_m = \sqrt{(2\pi \sigma^2)}\exp \left(\alpha_0 + \dfrac{u_m ^2}{2\sigma^2}\right).
    \label{eq: local para}
\end{equation}

\section{Spatial-time models} \label{sec: stmodel}
\subsection{\texorpdfstring{Integrating $\bm{\zeta}$}{Integrating zeta}}
Aiming at deriving spatial-time models with only $t$ and $\bm{x}$ as continuous variables, we need to eliminate the continuous variables $\xi \in \mathbb{R}$ and $\bm{\zeta} \in \mathbb{R}^{L}$ in $f_m$. For $\bm{\zeta}$, we define

\begin{equation}
	g_{m}(t, \bm{x}, \xi) = \int_{\mathbb{R}^{L}} f_{m}(t, \bm{x}, \xi, \bm{\zeta}) \text{d}\bm{\zeta}, \ \  h_{m}(t, \bm{x}, \xi) = \int_{\mathbb{R}^{L}}\left|\bm{\zeta}\right|^2 f_{m}(t, \bm{x}, \xi, \bm{\zeta}) \text{d}\bm{\zeta}
\end{equation}

We can see that the equilibrium state $g^{eq}_{m}$ of $g_{m}$ is exactly  the $\mathcal{E}_{tr, m}$ and $h^{eq}_{m} = L\theta g^{eq}_{m}$ with $\theta$ defined in equation \Ref{eq: temp}. Derived from equation \Ref{eq: DVD-BGK}, the equation of $g_m$ and $h_m$ are

\begin{equation}
        \left\{
	\begin{aligned}
	\partial_{t} g_m + \xi \bm{l}_m\cdot \nabla_{\bm{x}} g_m &=  \dfrac{1} {\tau}\left( \exp (\bm{\alpha}\cdot\bm{m}_m) - g_m \right)\\
	\partial_{t} h_m + \xi \bm{l}_m\cdot \nabla_{\bm{x}} h_m &=  \dfrac{1}{\tau}\left(L\theta \exp (\bm{\alpha}\cdot\bm{m}_m) - h_m \right)
	\end{aligned}
	\right.
\end{equation}

for $m = 1, 2, ..., N$. 

\subsection{Hyperbolic Quadrature Method of Moments (HyQMOM)}
Then, we treat $\xi$ with the hyperbolic quadrature method of moments (HyQMOM) \cite{fox22, Laboureur21} , which is one kind of methods of moments.  

Firstly, different from the conventional definition of moments, we define the moments of $g_m$ and $h_m$ as

\begin{equation}
    M_{m, k}^{[g]}(t, \bm{x}) = \int_{\mathbb{R}}g_{m}(t, \bm{x}, \xi)\xi^{k}\left|\xi\right|^{D-1}\text{d}\xi, \ \ 	M_{m, k}^{[h]}(t, \bm{x}) = \int_{\mathbb{R}}h_{m}(t, \bm{x}, \xi)\xi^{k}\left|\xi\right|^{D-1}\text{d}\xi
\end{equation}

for $m = 1, 2, ..., N$ and $k\in \mathbb{N}$. The new definitions have taken the weight function $\left|\xi\right|^{D-1}$ into consideration. 


Then, we integrate the BGK-DVDM equation \Ref{eq: DVD-BGK} to get the evolution equations for the moments $M_{m, k}^{[g]}$ and $M_{m, k}^{[h]}$

\begin{equation}
	\left\{
	\begin{aligned}
		\partial_{t}M_{m, k}^{[g]} + \bm{l}_{m}\cdot \nabla_{\bm{x}} M_{m, k+1}^{[g]}&= \dfrac{1}{\tau}(\rho_{m}\Delta_{k}(u_m, \sigma^2) - M_{m, k}^{[g]})\\
		\partial_{t}M_{m, k}^{[h]} + \bm{l}_{m}\cdot \nabla_{\bm{x}} M_{m, k+1}^{[h]}&= \dfrac{1}{\tau}(L\theta \rho_{m}\Delta_{k}(u_m, \sigma^2) - M_{m, k}^{[h]})
	\end{aligned}
	\right.
\end{equation}

for $k\in \mathbb{N}$. 

In the above evolution equations, $\Delta_{k}(u_m, \sigma^2):=\displaystyle \int_{\mathbb{R}}\mathcal{N}(u_m, \sigma^2)\xi^{k}\left|\xi\right|^{D-1}\text{d}\xi$ is the $k$-th moment of the normalized Gaussian distribution centered at $u_m$ with a variance $\sigma^2$, also taking the weight function $\left|\xi\right|^{D-1}$ into consideration directly. 

Then, we apply the HyQMOM method, which states that the $g_m(\xi)$ and $h_m(\xi)$ can be treated as a weighted sum of $2p+1$ Dirac delta functions, and $p$ depends on the number of moments we use naturally \cite{fox22}. 

\begin{equation}
    \phi_m (\xi)  = \sum_{\alpha = 1}^{2p+1} \omega_{m, \alpha}^{[\phi]} \delta(\xi - \lambda_{m, \alpha}^{[\phi]}),\: \text{for}\  \phi = g\  \text{or}\  h. 
    \label{eq: hy}
\end{equation}

Here the methods of getting the abscissas $\{\lambda_{m, \alpha}^{[\phi]}\}$ and weights $\{\omega_{m, \alpha}^{[\phi]}\}$ of $\phi$ will be talked about in Section \Ref{sec: cal}. In this way, the moment with order $2p+1$ can be approximated and expressed in terms of $\{\lambda_{m, \alpha}^{[\phi]}\}$ and $\{\omega_{m, \alpha}^{[\phi]}\}$, which means that the higher order moment could be expressed in terms of lower order moments, satisfying the closure of moments. 

\begin{equation}
    M_{m, 2p+1}^{[\phi]} = \sum_{\alpha = 1}^{2p+1} \omega_{m, \alpha}^{[\phi]} (\lambda_{m, \alpha}^{[\phi]})^{2p+1},\: \text{for}\  \phi = g\  \text{or}\  h. 
\end{equation}

Then, the macroscopic quantities can be calculated using the ansatz \Ref{eq: hy}:

\begin{align}
    \rho &= s\sum_{m, \alpha}\omega_{m, \alpha}^{[g]} \notag, \\
     \rho \bm{U} &= 
     s\sum_{m, \alpha}\bm{l}_{m} \omega_{m, \alpha}^{[g]} \lambda_{m, \alpha}^{[g]}, \\
    \rho E &= \dfrac{s}{2}
    \sum_{m, \alpha} \omega_{m, \alpha}^{[g]} (\lambda_{m, \alpha}^{[g]})^2 +  \omega_{m, \alpha}^{[h]}\notag.
\end{align}

This model is denoted as DVD-HyQMOM.

\section{Boundary conditions} \label{sec: bound}
Let $\Omega \in \mathbb{R}^{D}$ be the computational domain, and the $\bm{n}$ be the outside normalized vector of the boundary $\partial \Omega$. 

In this paper, the first kind of boundary conditions we adopt is Neumann boundary condition $\frac{\partial \phi}{\partial \bm{n}} = 0$, which means that the function $\phi$ keep unchanged along the $\bm{n}$ direction, and the values on the boundary cells are extended constantly along $\bm{n}$.

The second one is the solid wall conditions, which indicates the boundary distribution $f(t, \bm{x}_{w}, \bm{\xi}, \bm{\zeta})$ for reflecting particles should be determined by the distribution of outgoing particles. One specific boundary condition is the diffuse-scattering law \cite{Valougeorgis05}, assuming that the distribution of reflecting particles obey Maxwell distribution
\begin{equation}
	f(t, \bm{x}_{w}, \bm{\xi}, \bm{\zeta}) = \sqrt{\dfrac{2\pi}{\theta_{w}}}\dfrac{j(t, \bm{x}_{w})}{(2\pi \theta_{w})^{(D+L)/2}}\exp\left(-\dfrac{|\bm{\xi} - \bm{U}_{w}|^2 + |\bm{\zeta}|^2}{2\theta_{w}}\right),\quad \bm{\xi}\cdot \bm{n} < 0,
\end{equation}

in which $\bm{U}_{w}$ is the boundary velocity at $\bm{x}_{w}$, and $\theta_w$ is the boundary temperature at $\bm{x}_{w}$. The $j(t, \bm{x}_{w})$ is the outgoing mass flux
\begin{equation*}
    j(t, \bm{x}_{w}) = \int_{\mathbb{R}}\int_{\bm{\xi}\cdot \bm{n} >0} \bm{\xi}\cdot \bm{n} f(t, \bm{x}_{w}, \bm{\xi}, \bm{\zeta}) \text{d}\bm{\xi}\text{d}\bm{\zeta}.
\end{equation*}

This mass flux ensures that all particles are reflected, thus satisfying the conservation law with no particle loss or tunneling.

For the DVD-HyQMOM model, when using the diffuse-scattering law, the corresponding distributions of the reflected particle at the boundary can be written as

\begin{align}
    \begin{array}{lll}
    g_{m}(t, \bm{x}_w, \xi) &= \sqrt{\dfrac{2\pi}{\theta_{w}}}j(t, \bm{x}_{w})\mathcal{E}_{tr, m}[(1, \bm{U}_{w}, \theta_{w})], & \bm{\xi}\cdot\bm{n}< 0,\\
    h_{m}(t, \bm{x}_w, \xi) &= L\theta_{w}g_{m}(t, \bm{x}_w, \xi),& \bm{\xi}\cdot\bm{n}< 0,
    \end{array}
\end{align}

with
\begin{align*}
    j(t, \bm{x}_{w}) &= s\sum_{m = 1}^{N}\bm{l}_{m}\cdot \bm{n}\int_{\xi\bm{l}_{m}\cdot \bm{n} > 0}\xi\sum_{\alpha = 1}^{2p+1}\omega_{m, \alpha}^{[g]}\delta(\xi - \lambda_{m, \alpha}^{[g]})\text{d}\xi\\
     &= s\sum_{m = 1}^{N} \max\{\bm{l}_{m}\cdot \bm{n}, 0\}\sum_{\alpha = 1}^{2p+1}\omega_{m, \alpha}^{[g]}\max\{\lambda_{m, \alpha}^{[g]}, 0\} +  \min\{\bm{l}_{m}\cdot \bm{n}, 0\}\sum_{\alpha = 1}^{2p+1}\omega_{m, \alpha}^{[g]}\min\{\lambda_{m, \alpha}^{[g]}, 0\}
\end{align*}

and $\mathcal{E}_{tr, m}[(1, \bm{U}_{w}, \theta_{w})]$ the discrete equilibrium defined in equation \ref{eq: local eq} with density 1, velocity $\bm{U}_{w}$, and temperature $ \theta_{w}$.

\section{Algorithms} \label{sec: alg}
\subsection{Numerical schemes}
In this part, the numerical scheme of DVD-HyQMOM is presented. 
The moments $M_{m, k, i, j}^{[g]}$ and $M_{m, k, i, j}^{[h]}$ can be approximated by the 2-D upwind scheme \cite{fox13}
\begin{align}
	M_{m, k, i, j}^{[g], n+1} &= M_{m, k, i, j}^{[g],n} - \dfrac{\Delta t}{\Delta x}\bm{l}_{m}\cdot \bm{e}_1\left(\mathcal{F}_{m, k, i+1/2, j}^{[g],n} - \mathcal{F}_{m, k, i-1/2, j}^{[g],n}\right)\notag \\
	&-\dfrac{\Delta t}{\Delta y}\bm{l}_{m}\cdot \bm{e}_2\left(\mathcal{F}_{m, k, i, j+1/2}^{[g],n} - \mathcal{F}_{m, k, i, j-1/2}^{[g],n}\right) + \dfrac{\Delta t}{\tau}(M_{\mathcal{E}_m, k, i, j}^{[g],n} - M_{m, k, i, j}^{[g],n+1})
\end{align}

and

\begin{align}
	M_{m, k, i, j}^{[h], n+1} &= M_{m, k, i, j}^{[h],n} - \dfrac{\Delta t}{\Delta x}\bm{l}_{m}\cdot \bm{e}_1\left(\mathcal{F}_{m, k, i+1/2, j}^{[h],n} - \mathcal{F}_{m, k, i-1/2, j}^{[h],n}\right)\notag \\
	&-\dfrac{\Delta t}{\Delta y}\bm{l}_{m}\cdot \bm{e}_2\left(\mathcal{F}_{m, k, i, j+1/2}^{[h],n} - \mathcal{F}_{m, k, i, j-1/2}^{[h],n}\right) + \dfrac{\Delta t}{\tau}(M_{\mathcal{E}_m, k, i, j}^{[h],n} - M_{m, k, i, j}^{[h],n+1})
\end{align}

Here the moments $M_{\mathcal{E}_m, k, i, j}^{[g],n}$ and $M_{\mathcal{E}_m, k, i, j}^{[h],n}$ correspond to the equilibrium state, which can be expressed as 

\begin{equation*}
    M_{\mathcal{E}_m, k, i, j}^{[g],n} = \rho_{m, i, j}^{n} \Delta_k\left(u_{m, i, j}^n, (\sigma^2)_{i, j}^n\right)
\end{equation*}

and 

\begin{equation*}
    M_{\mathcal{E}_m, k, i, j}^{[h],n} = L \theta \rho_{m, i, j}^{n} \Delta_k\left(u_{m, i, j}^n, (\sigma^2)_{i, j}^n\right)
\end{equation*}

in which, $\Delta_k\left(u_{m, i, j}^n, (\sigma^2)_{i, j}^n\right)$ has already been defined in Section \ref{sec: dvdm}, which is the $k$-th moment of the normalized Gaussian distribution, and the specific expression can also be reflected in equation \ref{eq: local eq}. The equilibrium state parameters $\rho_{m, i, j}, u_{m, i, j}, (\sigma^2)_{i, j}^n$ could be obtained by minimizing the convex function $J(\mathbb{\alpha})$, and the explicit expressions of these parameters have already been reflected in equation \ref{eq: local para}.

The kinetic-based fluxes are

\begin{equation}
	\mathcal{F}_{m, k, i+1/2, j}^{[g], n} =\left\{
	\begin{aligned}
		&\int_{0}^{\infty}g_{m, i, j}^{ n}\xi^{k+1}\left|\xi\right|^{D-1}\text{d}\xi  + \int_{-\infty}^{0}g_{m, i+1, j}^{ n}\xi^{k+1}\left|\xi\right|^{D-1}\text{d}\xi ,\ \ &\bm{l}_m\cdot\bm{e}_1 > 0 \\
		& \int_{0}^{\infty}g_{m, i+1, j}^{ n}\xi^{k+1}\left|\xi\right|^{D-1}\text{d}\xi  + \int_{-\infty}^{0}g_{m, i, j}^{ n}\xi^{k+1}\left|\xi\right|^{D-1}\text{d}\xi,\ \ &\bm{l}_m\cdot\bm{e}_1 < 0
	\end{aligned}
	\right.
\end{equation}

and 

\begin{equation}
	\mathcal{F}_{m, k, i, j+1/2}^{[g], n} =\left\{
	\begin{aligned}
		&\int_{0}^{\infty}g_{m, i, j}^{ n}\xi^{k+1}\left|\xi\right|^{D-1}\text{d}\xi  + \int_{-\infty}^{0}g_{m, i, j+1}^{ n}\xi^{k+1}\left|\xi\right|^{D-1}\text{d}\xi  ,\ \ &\bm{l}_m\cdot\bm{e}_2 > 0 \\
		& \int_{0}^{\infty}g_{m, i, j+1}^{ n}\xi^{k+1}\left|\xi\right|^{D-1}\text{d}\xi  + \int_{-\infty}^{0}g_{m, i, j}^{ n}\xi^{k+1}\left|\xi\right|^{D-1}\text{d}\xi ,\ \ &\bm{l}_m\cdot\bm{e}_2 < 0
	\end{aligned}
	\right.
\end{equation}

and the fluxes for $h$ could be treated similarly.

The change in the moments due to collisions over a time step $\Delta t$ can be evaluated explicitly at each grid cell:
\begin{equation*}
	M^{*} = M_{\mathcal{E}_m} + (M- M_{\mathcal{E}_m})\exp(-\Delta t/\tau)
\end{equation*}

\subsection{How to get the abscissas and weights}

\subsubsection{Computation of abscissas using Chebyshev algorithm}
\label{sec: cal}
According to the HyQMOM closure and the global hyperbolicity \cite{fox22}, the characteristic polynomial $P_{2p+1}$ of the Jacobian matrix $\frac{D\bm{F}}{D\bm{M}}$ of system $\partial_{t}\bm{M} + \partial_{x}\bm{F}(\bm{M}) = 0$ can be written as $P_{2p+1} = Q_{p}R_{p+1}$ if and only if the coefficients $\alpha_p, \beta_{p}$ and $a_p$ satisfy the following condition

\begin{equation}
    a_p = \alpha_p = \dfrac{1}{p}\sum_{t=0}^{p-1} a_{t}, \quad \beta_p = \dfrac{2p+1}{p} b_p
\end{equation}

Here the recursion coefficients $a_{t}\ (t = 0, 1, ..., p-1)$ and $b_{t}\ (t = 0, 1, ..., p)$ can be obtained from the first $2s+1$ order moments $M_{m, k, i, j}^{[\phi], n}\ (k = 0, 1, ..., 2p)$ through the Chebyshev algorithm \cite{cheby1, cheby2}.  

Then the roots of the two polynomials $Q_{s}$ and $R_{s+1}$ can be obtained by calculating the eigenvalues of the following two Jacobi matrices \cite{fox22},

    \begin{equation}
        \textbf{J}_{p} = \begin{pmatrix}
            a_{0}&\sqrt{b_1}  &  &  & \\
            \sqrt{b_1}&  a_{1}&\sqrt{b_2} &  &  \\
            & \ddots &\ddots  &\ddots  & \\
            &&\sqrt{b_{p-2}}&a_{p-2}&\sqrt{b_{p-1}} \\
            &  &  & \sqrt{b_{p-1}} &a_{p-1}  
        \end{pmatrix}
    \end{equation}

 and 

    \begin{equation}
        \textbf{K}_{p+1} = \begin{pmatrix}
            a_{0}&\sqrt{b_1}  &  &  &  \\
            \sqrt{b_1}&  a_{1}&\sqrt{b_2} &   & \\
            &\ddots & \ddots& \ddots & \\
            &&\sqrt{b_{p-1}}&a_{p-1}&\sqrt{\beta_{p}} \\
            &  &  &\sqrt{\beta_{p}}  &\alpha_{p}
        \end{pmatrix}
    \end{equation}.

corresponding to the $2p+1$ abscissas, denoted as $\lambda_{m, i, j, \alpha}^{[\phi], n}\ (\alpha = 0, 1, ..., 2p)$.

\subsubsection{Computation of weights}
Here we give two methods to compute the weights for the abscissas. Since their performances are almost the same, we only use the first method in the following numerical examples.

\textbf{a.}
The first method is straightforward, which is to directly invert the matrix composed of different orders of abscissas, or in other words, solving the following linear equation
\begin{equation*}
    \overline{\bm{\Lambda}} \bm{\omega} = \bm{m}
\end{equation*}

in which 
 \begin{align*}
\overline{\bm{\Lambda}} = M_{m, 0, i, j}^{[\phi], n}\begin{pmatrix}
        1&1  & 1 &1  &\cdots&1  \\
        \lambda_{m, i, j, 0}^{[\phi], n}&  \lambda_{m, i, j, 1}^{[\phi], n}&\lambda_{m, i, j, 2}^{[\phi], n} & \lambda_{m, i, j, 3}^{[\phi], n}  &\cdots&\lambda_{m, i, j, 2p}^{[\phi], n} \\
        (\lambda_{m, i, j, 0}^{[\phi], n})^2&(\lambda_{m, i, j, 1}^{[\phi], n})^2 & (\lambda_{m, i, j, 2}^{[\phi], n})^2& \lambda_{m, i, j, 3}^{[\phi], n})^2& \cdots&(\lambda_{m, i, j, 2p}^{[\phi], n})^2\\
        (\lambda_{m, i, j, 0}^{[\phi], n})^3&(\lambda_{m, i, j, 1}^{[\phi], n})^3&(\lambda_{m, i, j, 2}^{[\phi], n})^3&(\lambda_{m, i, j, 3}^{[\phi], n})^3&\cdots&(\lambda_{m, i, j, 2p}^{[\phi], n})^3\\
        \vdots&\vdots  & \vdots &  \vdots &\ddots&\vdots\\
        (\lambda_{m, i, j, 0}^{[\phi], n})^{2p} & (\lambda_{m, i, j, 1}^{[\phi], n})^{2p} & (\lambda_{m, i, j, 2}^{[\phi], n})^{2p} & (\lambda_{m, i, j, 3}^{[\phi], n})^{2p}& \cdots & (\lambda_{m, i, j, 2p}^{[\phi], n})^{2p}
    \end{pmatrix},
    \end{align*}
    \begin{align*}
     \bm{\omega} =\begin{pmatrix}
        \omega_{m, i, j, 0}^{[\phi], n}\\
        \omega_{m, i, j, 1}^{[\phi], n}\\
        \omega_{m, i, j, 2}^{[\phi], n}\\
        \omega_{m, i, j, 3}^{[\phi], n}\\
        \cdots\\
        \omega_{m, i, j, 2p}^{[\phi], n}
    \end{pmatrix}, \quad
    \bm{m} = \begin{pmatrix}
        M_{m, 0, i, j}^{[\phi], n}\\
        M_{m, 1, i, j}^{[\phi], n}\\
        M_{m, 2, i, j}^{[\phi], n}\\
        M_{m, 3, i, j}^{[\phi], n}\\
        \vdots\\
        M_{m, 2p, i, j}^{[\phi], n}\\
        \end{pmatrix}.
\end{align*}
 
In this matrix, $M_{m, k, i, j}^{[\phi], n}$ represents the $k$-th order moment of function $\phi$ ($\phi$ equals to $g$ or $h$, $k = 0, 1, ..., 2p$), and the $\lambda_{m, i, j, \alpha}^{[\phi], n}$ corresponds to the $\alpha$-th eigenvalue, which is to say the $\alpha$-th abscissa ($\alpha = 0, 1, ..., 2p$). Notice that the matrix is a Vandermonde matrix, so it is theoretically reversible.

\textbf{b.} 
Once we get the Jacobi matrices, the weights could be calculated directly from the eigenvectors using the Christoffel-Darboux identity \cite{qmom1997}

\begin{equation}
    \omega_{m, i, j, \alpha}^{[\phi], n} = M_{m, 0, i, j}^{[\phi], n} (v_{m, i, j, \alpha 1}^{[\phi], n})^2
\end{equation}

Here the $v_{m, i, j, \alpha 1}^{[\phi], n}$ is the first component of the $\alpha$-th eigenvector $\bm{v}_{m, i, j, \alpha}^{[\phi], n}$, corresponding to the $\alpha$-th eigenvalue. One major advantage of this method is that it is not constrained by dimensional limitations, such that it can be effectively utilized within the QMOM method \cite{qmom1997} and the GQMOM method \cite{gqmom}.


\section{Numerical results} \label{sec: ex}
In this section, we present some numerical results, mainly focusing on the discretizations of DVD-HyQMOM model, also including comparisons with other DVDM models, such as DVD-EQMOM model and DVD-GQMOM model. The tests only involve planar flows $(D = 2)$.
\subsection{1-D Riemann problems}
Firstly, we consider 1-D Riemann problems, assuming no internal degrees of freedom $(L = 0)$. Initial conditions of the Riemann problems we consider are as follows \cite{fox2008}:

$$ \rho(0, x) = \left\{
\begin{aligned}
 &3.093, &x<0, \\
 &1, &x>0, 
\end{aligned}
\quad \bm{U}(0, x) = \bm{0}, \quad \theta(0, x) = 1,
\right.
$$
where $\rho$, $\bm{U}$, $\theta$ represents density, velocity and temperature separately in classical fluid problems, consistent with the definition of symbols in theoretical part. For the boundary condition, we apply Neumann boundary condition $\frac{\partial f}{\partial \bm{n}} = 0$, which means the values on the boundary cells are extended constantly along the outward-facing unit normal vector $\bm{n}$. 

The 1-D physical domain $\left[-0.5, 0.5\right]$ is divided into 200 uniform cells. We adopt two limiting approximations, one for the continuum regime (infinitely fast collision limit with $\tau = 0$), and the other for the free-molecular regime (no collision limit with $\tau = \infty$), both of which are considered in our numerical cases. In practical calculations, the continuum regime is characterised with $\tau = 10^{-4}$, and the free-molecular regime is characterised with $\tau = 10^{4}$. In all DVDM models used for 1-D Riemann problems, we set $N = 8$, and the corresponding directions are $\left\{\bm{l}_m = \left(\cos\frac{(2m-1)\pi}{16}, \sin\frac{(2m-1)\pi}{16}\right)^{T}\right\}^{8}_{m = 1}$. 

\begin{figure}[H]
    \centering
    \includegraphics[scale = 0.45]{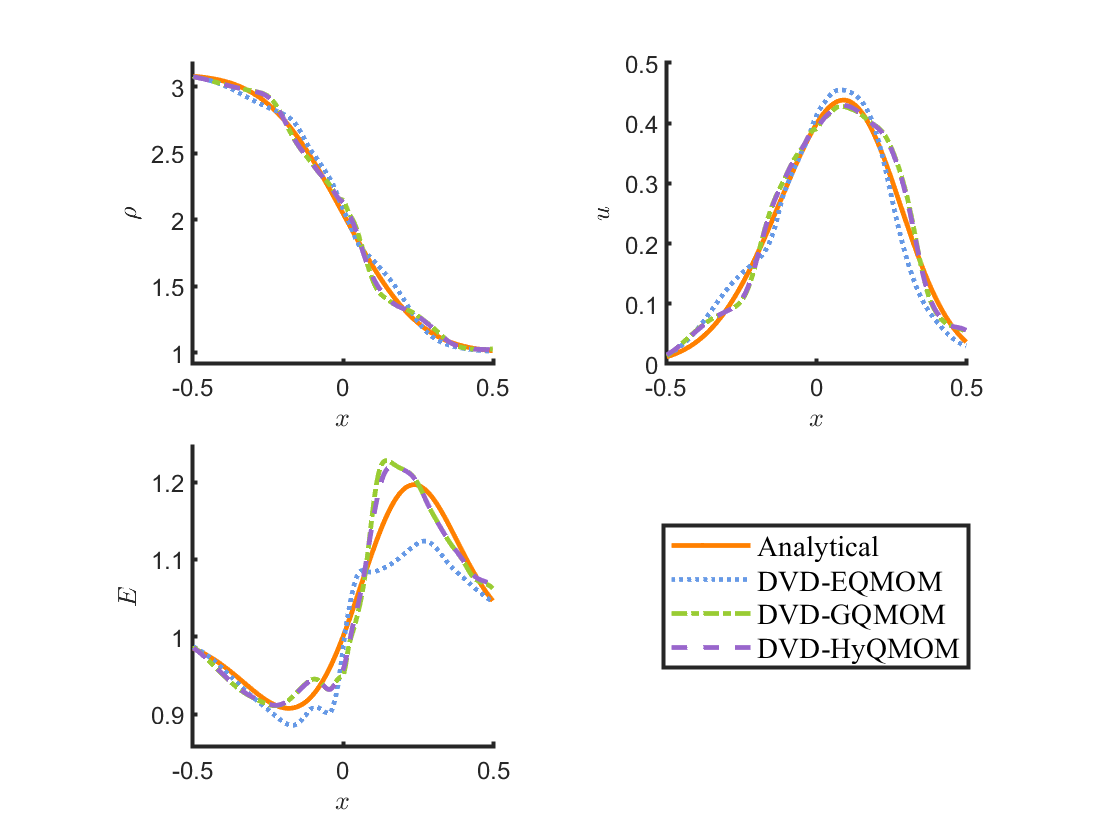}
    \caption{1-D Riemann problem with $\tau = 10^4$: profiles of density $\rho$, velocity $u$ and energy $E$ at $t = 0.2$. Orange, analytical solution. Blue, the DVD-EQMOM model. Green, the DVD-GQMOM model. Purple, the DVD-HyQMOM mode. Here the analytical result is the solution of the free-transport equation. In all models, we set $N = 8$ and the directions  $\left\{\bm{l}_m = \left(\cos\frac{(2m-1)\pi}{16}, \sin\frac{(2m-1)\pi}{16}\right)^{T}\right\}^{8}_{m = 1}$. We set $n = 2$ for the DVD-HyQMOM and the DVD-GQMOM, and $M = 2$ for the DVD-EQMOM.}
    \label{1dmethodcoless}
\end{figure}

Let us begin with the free-molecular regime. Figure \ref{1dmethodcoless} exhibits the spatial distributions of the macroscopic quantities $(\rho, u, E)$ at $t = 0.2$, including the simulated results of different DVDM models and the analytical result found in \cite{xu15}, solved from the Eular equations as well. 
For the DVD-HyQMOM, we set $n = 2$, which means the system has five moments and is simulated with five abscissas and five weights on each direction. It should be noted that this is the simplest case of HyQMOM method, using the least number of moments. For the DVD-GQMOM, we still choose $n = 2$ to match the previous DVD-HyQMOM model. For the DVD-EQMOM \cite{chen2}, we set $M = 2$. 

It is seen that the DVD-HyQMOM model is feasible, although there seems to be no much advantage in accuracy compared with the DVD-EQMOM model. It is worth noticing that, just as mentioned before, the DVD-HyQMOM used here is just the simplest case, and increasing the number of abscissas and weights is expected to make the results more accurate, which could be achieved by the HyQMOM method, but not by the others.

\begin{figure}[H]
    \centering
    \includegraphics[scale = 0.55]{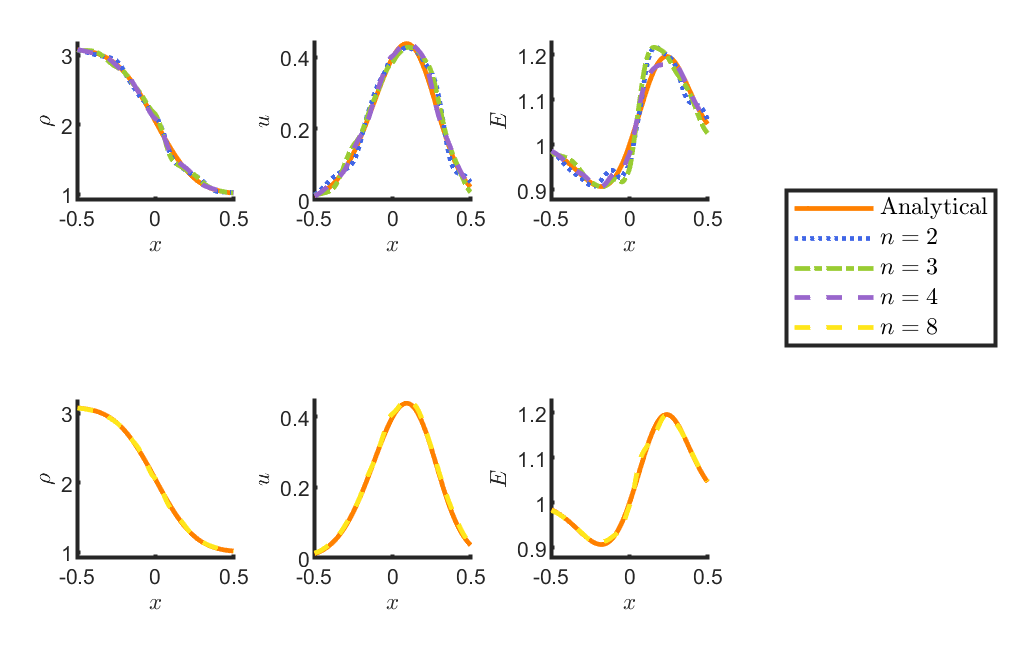}
    \caption{1-D Riemann problem with $\tau = 10^4$: profiles of density $\rho$, velocity $u$ and energy $E$ at $t = 0.2$. Orange, analytical solution. Blue, $n=2$. Green, $n = 3$. Purple, $n = 4$. Yellow, $n = 8$. In all cases, we set $N = 8$ and the directions  $\left\{\bm{l}_m = \left(\cos\frac{(2m-1)\pi}{16}, \sin\frac{(2m-1)\pi}{16}\right)^{T}\right\}^{8}_{m = 1}$, using the DVD-HyQMOM model.}
    \label{1dncoless}
\end{figure}

Therefore, we focus on analysing the results of the DVD-HyQMOM model with different $n$ next, verifying that the simulation results will get better as the number of abscissas and weights increase in this model. As can be seen from the first line in Figure \ref{1dncoless}, the DVD-HyQMOM simulation results for the classical fluid microscopic quantities agree with the analytical results better with increasing $n$. 

\begin{figure}[ht]
    \centering
    \begin{subfigure}[b]{0.45\textwidth}
        \centering
        \captionsetup{justification=raggedright, singlelinecheck=false, position=top, aboveskip=-1pt, belowskip=-1pt}
        \caption{}
        \includegraphics[width=\textwidth, height=0.8\textwidth]{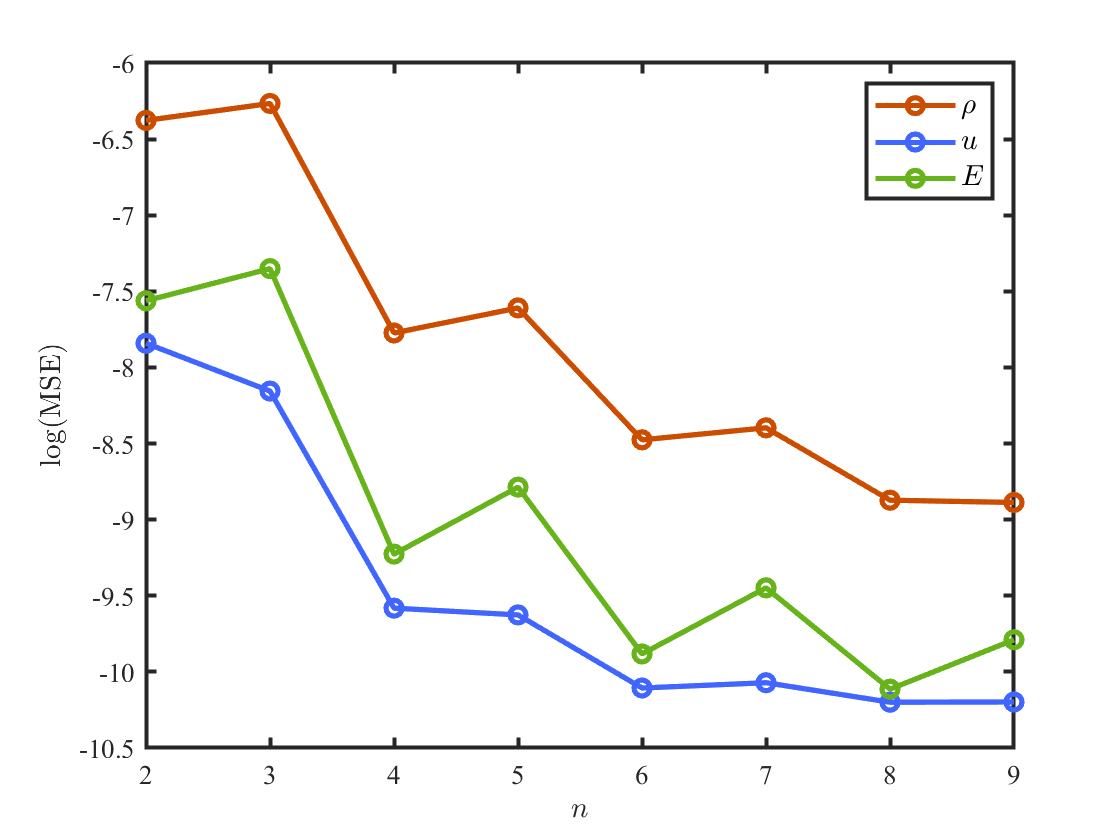}
        \label{1derror} 
    \end{subfigure}
    \begin{subfigure}[b]{0.45\textwidth}
        \centering
        \captionsetup{justification=raggedright, singlelinecheck=false, position=top, aboveskip=-1pt, belowskip=-1pt}
        \caption{}
        \includegraphics[width=1\textwidth, height=0.8\textwidth]{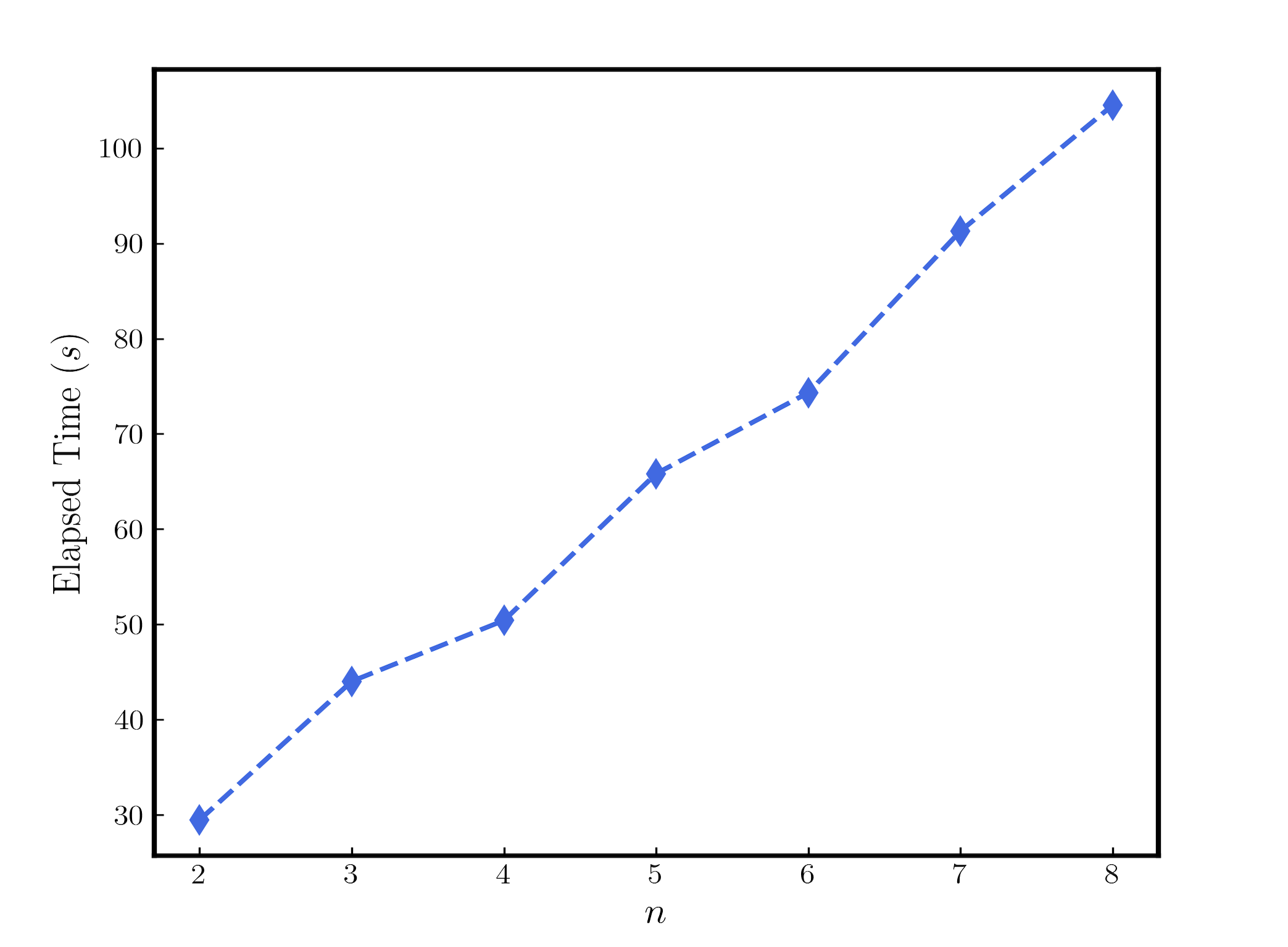} 
        \label{1dtime}
    \end{subfigure}
    
    \vskip\baselineskip
    \caption{1-D Riemann problem with $\tau = 10^4$ at $t = 0.2$: (a) The correlation between the Mean Squared Errors of classical fluid quantities and $n$. Orange, $\rho$. Green, $E$. Blue, $u$. (b) The correlation between the elapsed time and $n$.}
\end{figure}

To be more accurately and intuitively, the Mean Squared Error (MSE) of different cases with different $n$ are plotted in Figure \ref{1derror}. Here, we present the logarithmic values of the Mean Squared Error to more clearly illustrate the variation of the error. It can be observed that as $n$ increases, the error exhibits significant orders of magnitude differences, implying a dramatic effect of increasing $n$. Simultaneously, the results of the DVD-HyQMOM model simulation with $n = 8$ are plotted on the second line in Figure \ref{1dncoless}. It can be seen that these results agree with the analytical solutions extremely well, the effect of which is quite difficult to achieve using other momentum methods combined with DVD-model. 

To demonstrate the feasibility of the DVD-HyQMOM model, the elapsed time required for different $n$ is also presented in Figure \ref{1dtime}. 

\begin{figure}[H]
    \centering
    \includegraphics[scale = 0.45]{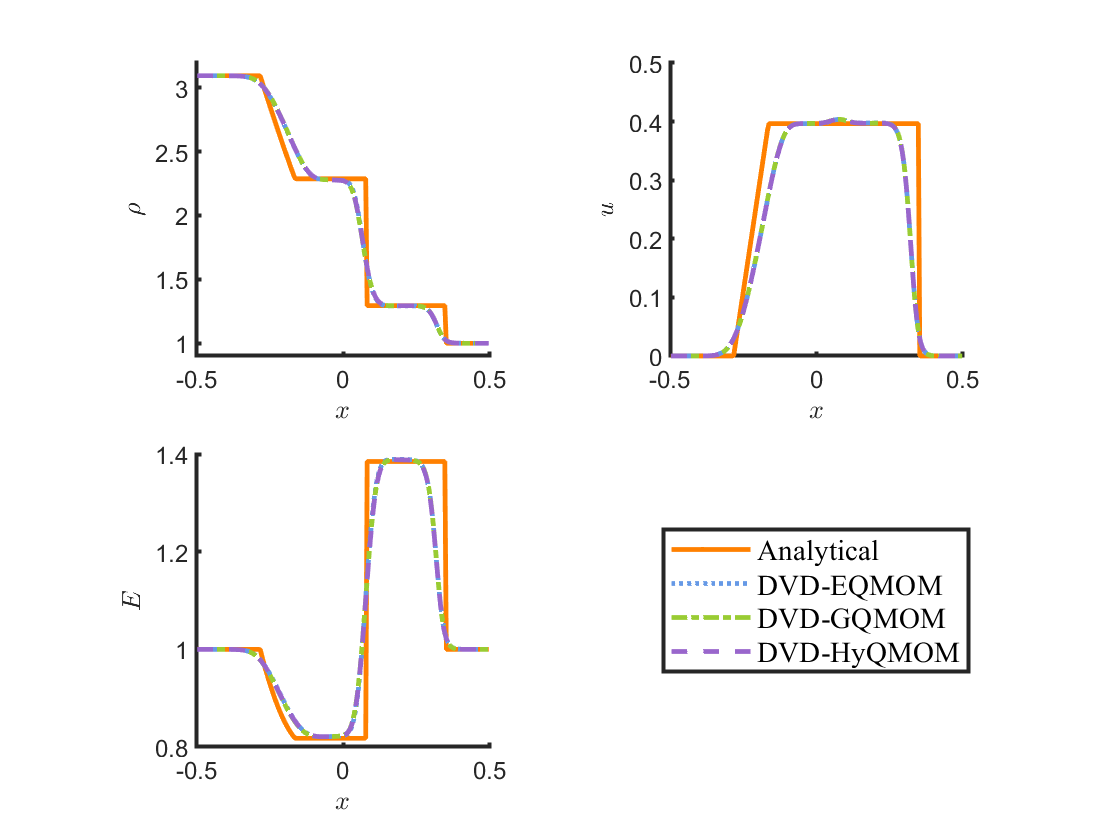}
    \caption{1-D Riemann problem with $\tau = 10^{-4}$: profiles of density $\rho$, velocity $u$ and energy $E$ at $t = 0.2$. Orange, analytical solution. Blue, the DVD-EQMOM model. Green, the DVD-GQMOM model. Purple, the DVD-HyQMOM model. Here the analytical result is the solution of the Euler equations \cite{xu15, 1dana0}. In all models, we set $N = 8$ and the directions  $\left\{\bm{l}_m = \left(\cos\frac{(2m-1)\pi}{16}, \sin\frac{(2m-1)\pi}{16}\right)^{T}\right\}^{8}_{m = 1}$. We set $n = 2$ for the DVD-HyQMOM and the DVD-GQMOM, and $M = 2$ for the DVD-EQMOM.}
    \label{1dmethodcol}
\end{figure}

Then we move on to the continuum regime. Same as the free-molecular regime, we compare the simulation results of different DVDM models for the classical fluid quantities, $\rho$, $u$, and $E$, which are shown in Figure \ref{1dmethodcol}. The conditions are also the same, that is, we set $n=2$ for the DVD-HyQMOM and the DVD-GQMOM, and $M=2$ for the DVD-EQMOM. Interestingly, there is no significant difference in the simulation results of the three moment methods at this time, different from the situation in the free-molecular regime. 

\begin{figure}[H]
    \centering
    \includegraphics[scale = 0.55]{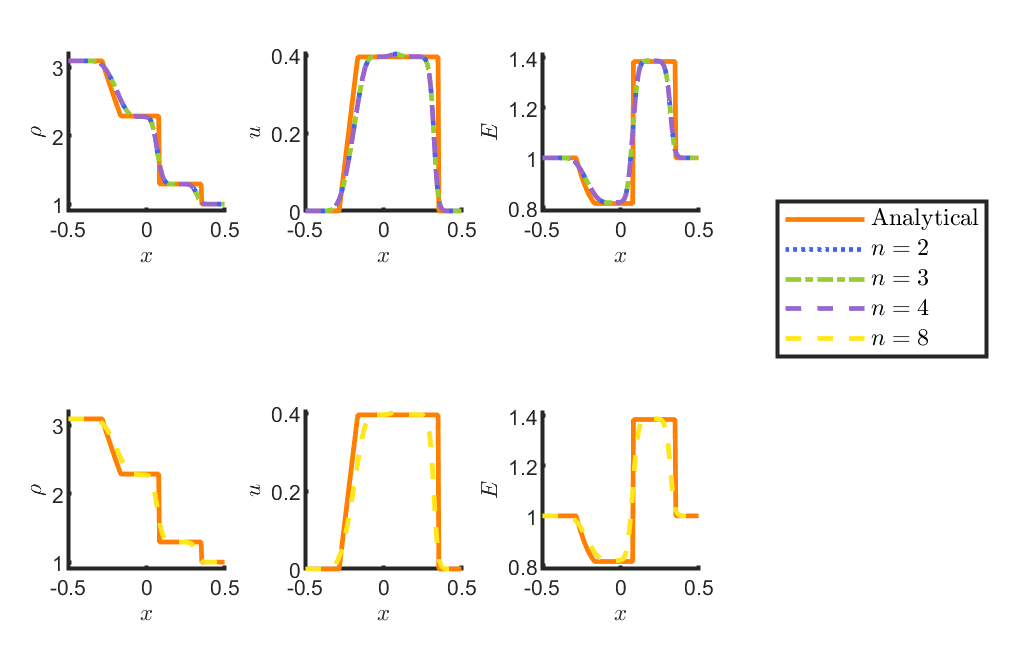}
    \caption{1-D Riemann problem with $\tau = 10^{-4}$: profiles of density $\rho$, velocity $u$ and energy $E$ at $t = 0.2$. Orange, analytical solution. Blue, $n=2$. Green, $n = 3$. Purple, $n = 4$. In all cases, we set $N = 8$ and the directions  $\left\{\bm{l}_m = \left(\cos\frac{(2m-1)\pi}{16}, \sin\frac{(2m-1)\pi}{16}\right)^{T}\right\}^{8}_{m = 1}$, using the DVD-HyQMOM model.}
    \label{1dncol}
\end{figure}

Meanwhile, increasing the number of $n$, which means increasing the number of abscissas, does not seem to bring much benefit in the problem of continuum regime as well. The contradistinction between the simulation results with different number of $n$ is shown in the first line of Figure \ref{1dncol}. We also conduct simulations under the condition of $n = 8$, as shown in the second line of Figure \ref{1dncol}. Nonetheless, the effect still has not been significantly improved.

\subsection{Couette flow}
For the Couette flow problem, the flow is restricted between two infinite parallel walls. Here we set the locations of the walls at $x = \pm 0.5 H$. The right wall at $x = 0.5 H$ moves with a constant velocity $\bm{v}_{w} = v_{w} \bm{e}_{y}$, on the other hand, the left wall moves at a speed of $-v_{w} \bm{e}_{y}$, which drive the fluid between them to a steady state, implying that the distribution $f$ maintains consistency on $y$ direction, and only depends on $x$. Here we set $H = 1$, $v_{w} = 0.1$, and the wall temperature $\theta_{w} = 2$. Initial conditions of of the Couette flow we consider are $\left(\rho, \bm{U}_0, \theta_0\right) = (1, \bm{0}, 2)$. 

For the DVD-HyQMOM simulation, we divide the domain between two walls $[0.5, -0.5]$ into 200 uniform cells, and set $N = 16$, with the corresponding velocity directions \\ $\left\{\bm{l}_m = \left(\cos\frac{(2m-1)\pi}{32}, \sin\frac{(2m-1)\pi}{32}\right)^{T}\right\}^{16}_{m = 1}$. Instead of the Neumann boundary condition, here the fluid is confined between the two parallel walls, so we should apply the diffuse-scattering law as the wall boundary condition. The Knudsen number Kn is an important representative parameter for the Couette flow, which is defined as the radio of the distance between two walls $H$ and the gas mean-free-path $\lambda$ \cite{wangcou13}:
\begin{equation*}
    \text{Kn} = \dfrac{\lambda}{D} = \dfrac{\tau}{D}\sqrt{\dfrac{\pi \theta_0}{2}}.
\end{equation*}

Based on the Knudsen number, gas flows can be classified as continuum ($\text{Kn} < 0.01$), slip ($0.01 < \text{Kn} < 0.1$), transition ($0.1 < \text{Kn} < 10$), and free-molecular ($\text{Kn} > 10$) flow regimes \cite{Bahukudumbi03}. Here the $\tau$ presents the characteristic collision time, same as that in the BGK equation, which indicates that we could tune the Couette flow regime by changing the number of $\tau$. Another useful parameter $\kappa$ is defined as $\kappa := \frac{\sqrt{\pi}}{2}\text{Kn}$. 

\begin{figure}[ht]
    \centering
    \begin{subfigure}[b]{0.455\textwidth}
        \centering
        \captionsetup{justification=raggedright, singlelinecheck=false, position=top, aboveskip=-1pt, belowskip=-1pt}
        \caption{}
        \includegraphics[width=\textwidth, height=0.9\textwidth]{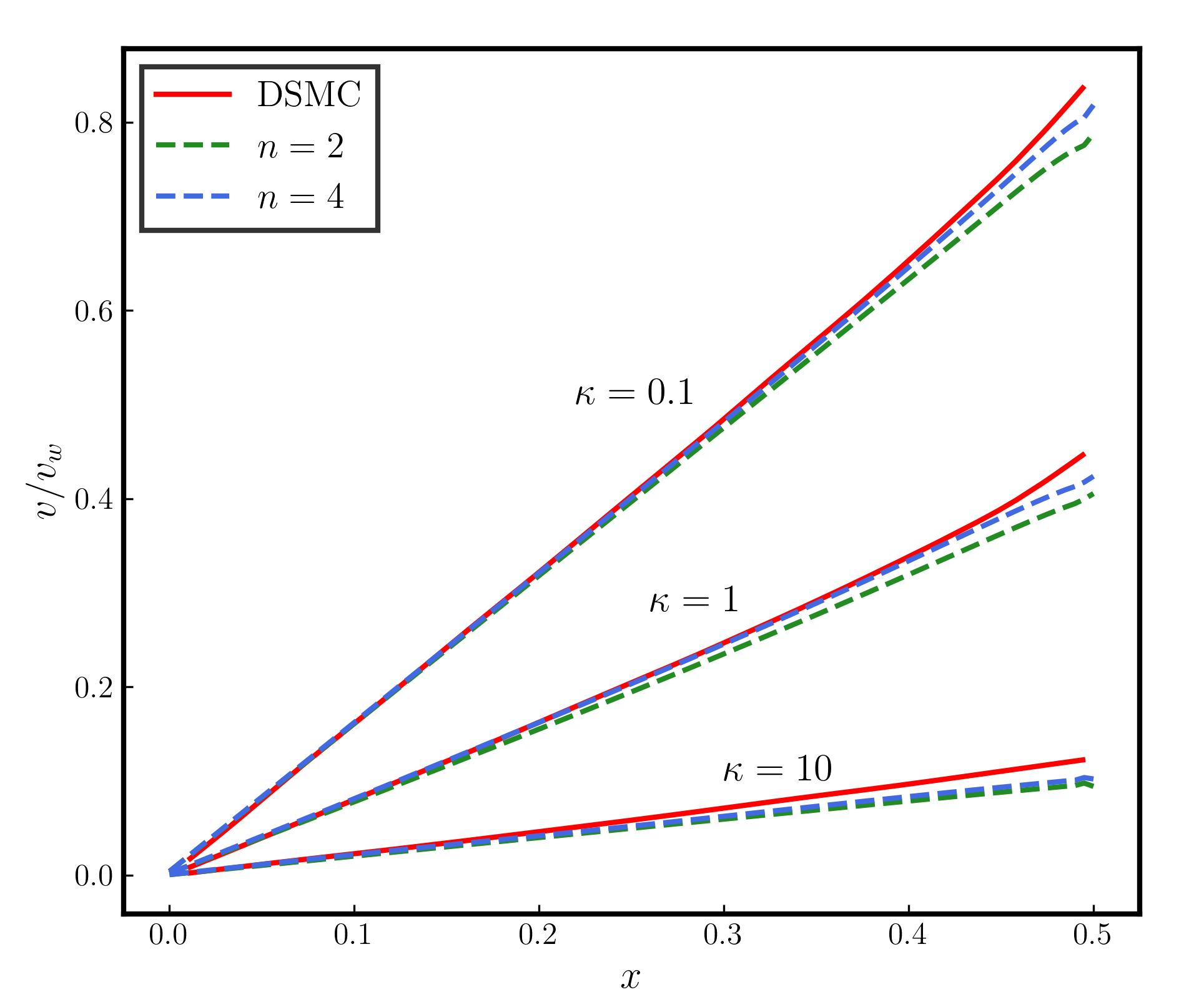}
        \label{coue_v}
    \end{subfigure}
    \begin{subfigure}[b]{0.453\textwidth}
        \centering
        \captionsetup{justification=raggedright, singlelinecheck=false, position=top, aboveskip=-1pt, belowskip=-1pt}
        \caption{}
        \includegraphics[width=\textwidth, height=0.9\textwidth]{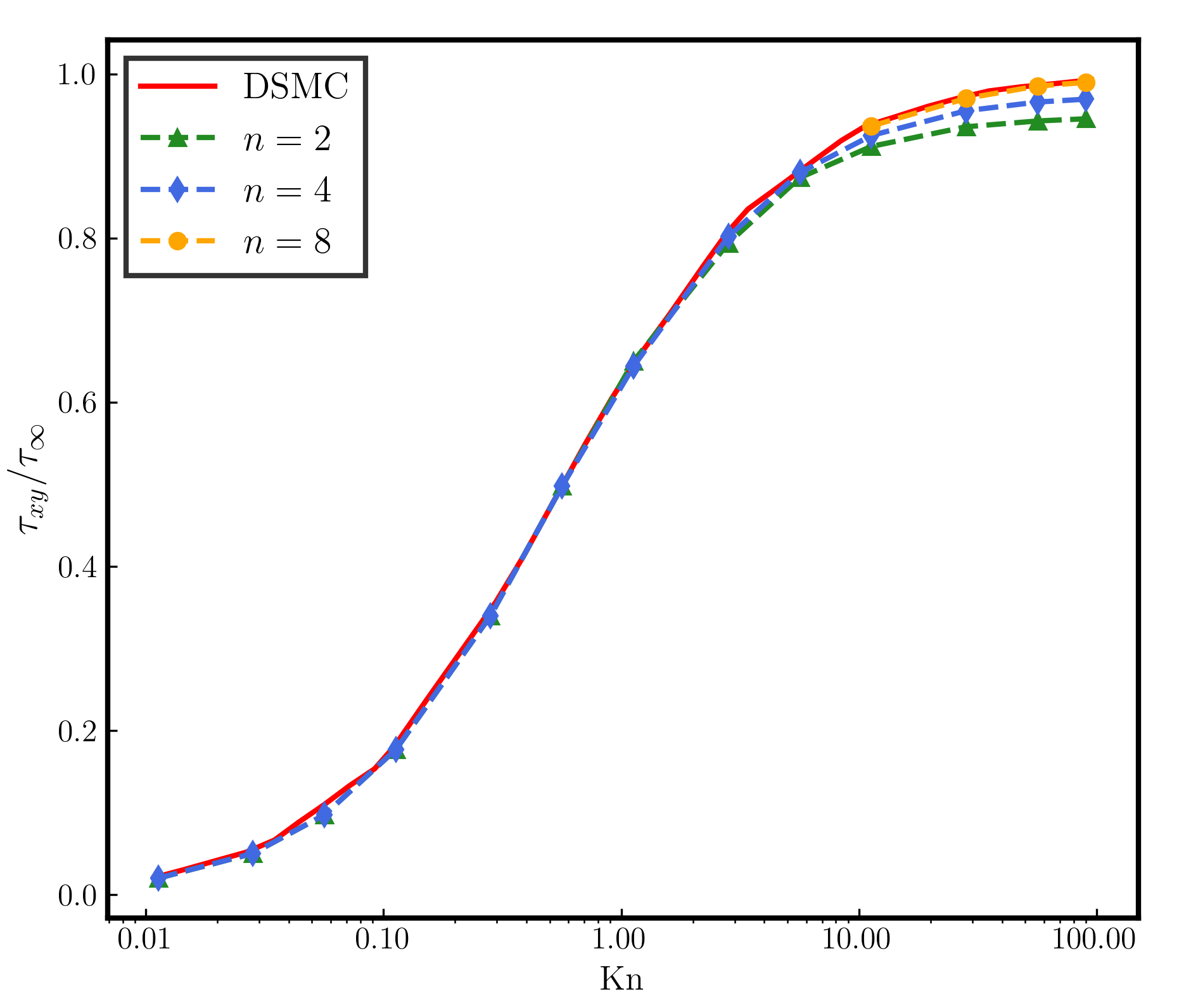} 
        \label{coue_tau}
    \end{subfigure}
    \vskip\baselineskip
    \caption{Couette flow: (a) Steady-State vertical velocity profiles with $\kappa = 0.1, 1, \text{and} 10$. Red, DSMC data as reference. Green, $n = 2$. Blue, $n = 4$. (b) Shear stress for different values of Kn. Red, DSMC data as reference. Green, $n = 2$. Blue, $n = 4$. Orange, $n = 8$. In all cases, we set $N = 16$ ad the directions $\left\{\bm{l}_m = \left(\cos\frac{(2m-1)\pi}{32}, \sin\frac{(2m-1)\pi}{32}\right)^{T}\right\}^{16}_{m = 1}$, using the DVD-HyQMOM model.}
\end{figure}

Figure \ref{coue_v} shows the steady-state velocity profiles on the positive domain $[0, 0.5]$ for different numbers of $\kappa$. The velocity is normalized by the constant velocity of the wall $v_{w}$. Figure \ref{coue_tau} further presents the shear stress $\tau_{xy}$ which is defined by:
\begin{align*}
    \tau_{xy} &= \left< (\xi_{x} - u)(\xi_{y} - v)f(\bm{\xi}) \right> = \left<\xi_{x}\xi_{y}\right>- \rho u v \\
     &= s\sum_{m = 1}^{N} \int_{\mathbb{R}}\int_{\mathbb{R}^{L}} (\xi \cos \theta_{m})(\xi \sin \theta_{m})f_{m} \left|\xi\right|^{D-1} \text{d}\bm{\zeta}\text{d}\xi - \rho u v
\end{align*}

for Kn ranging from 0.01 to 100. The shear stress is normalized by the free-molecular stress $\tau_{\infty} = -\rho v_{w}\sqrt{\frac{2\theta_0}{\pi}}$. In all cases, we show the DSMC results as reference \cite{wangcou13}. We also compare the results of the DVD-HyQMOM model with different $n$ in both the steady-state velocity and the shear stress cases. 

It is seen that the DVD-HyQMOM can reproduce the velocity profiles close to the reference data, especially when $\kappa$ is not too large. A minor drawback is that the treatment at the boundaries lacks sufficient precision and detail. For the shear stress, Figure \ref{coue_tau} shows that the few-nodes DVD-HyQMOM model has more significant errors at large Kn, and this issue is also evident in other moment methods. However, as $n$ increases, which means the number of abscissas increases, the velocity and shear stress profiles will be much closer to the reference data, bringing very effective correction effect, which is consistent with the conclusion given in the 1-D Riemann problem. 

\subsection{2-D Riemann problems}
In this part, we move to 2-D Riemann Problems, with the following initial conditions \cite{liu98}:
\begin{equation*}
    (\rho, u, v, p) = \left\{ 
  \begin{array}{lll}
    (\rho_1, u_1, v_1, p_1) = (1.1, 0, 0, 1.1)& x > 0, & y>0\\
     (\rho_2, u_2, v_2, p_2) = (0.5065, 0.8939, 0, 0.35)& x\le 0, & y>0\\
     (\rho_3, u_3, v_3, p_3) = (1.1, 0.8939, 0.8939, 1.1)& x\le 0, & y\le 0\\
     (\rho_4, u_4, v_4, p_4) = (0.5065, 0, 0.8939, 0.35)& x> 0, & y\le 0\\
  \end{array}
\right.
\end{equation*}

Different from the one-dimensional problems, here the internal degrees of freedom $L$ should also be taken into consideration. Therefore, we set $L = 3$, then the specific heat ratio $\gamma = (2 + D + L)/(D + L) = 1.4$. 

The 2-D physical domain is $[-0.5, 0.5]^{2}$, which is divided into $200\times 200$ uniform cells. Same as the 1-D Riemann problem, the Neumann boundary condition is applied to 2-D Riemann problem. We still consider both the free-molecular regime, characterized with $\tau = 10^{4}$, and the continuum regime, characterized with $\tau = 10^{-4}$.

\begin{figure}[ht]
    \centering
    \begin{subfigure}[b]{0.32\textwidth}
        \centering
        \captionsetup{justification=raggedright, singlelinecheck=false, position=top, aboveskip=-1pt, belowskip=-1pt}
        \caption*{(a)}

        \includegraphics[width=\textwidth, height=0.9\textwidth]{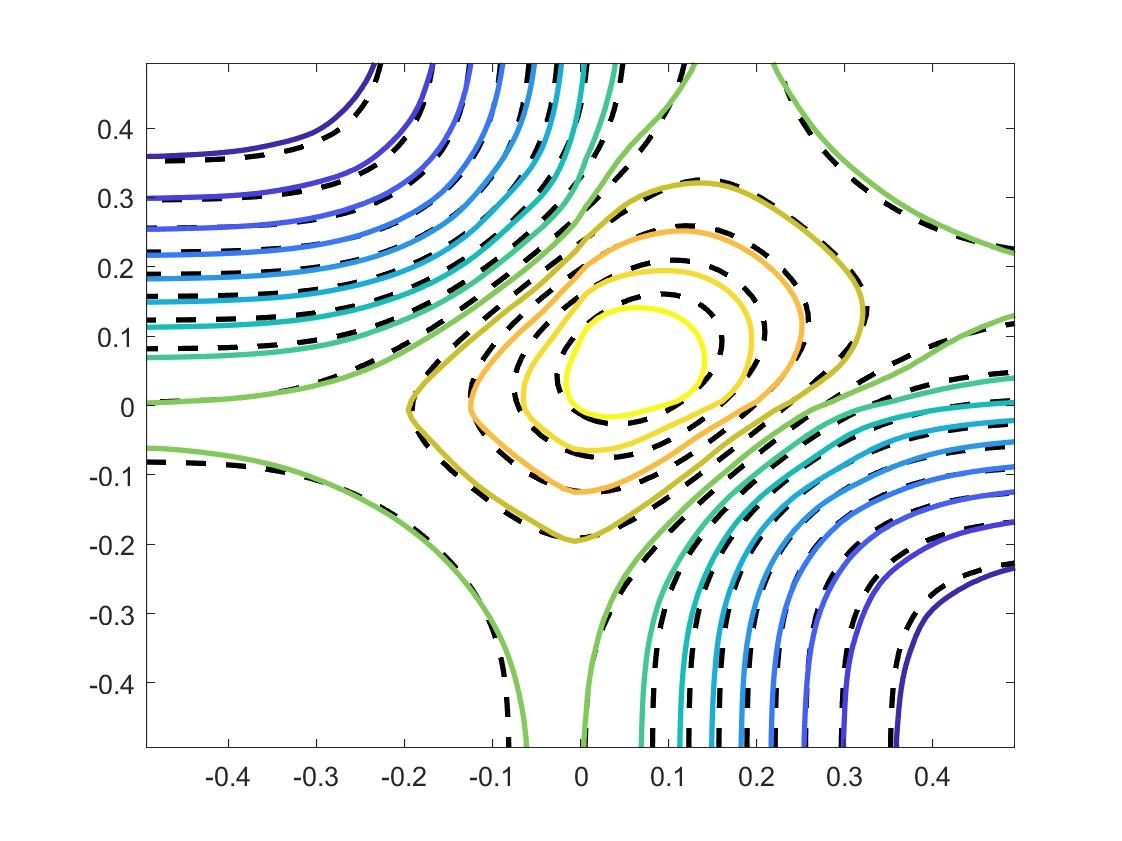}
    \end{subfigure}
     \hfill
    \begin{subfigure}[b]{0.32\textwidth}
        \centering
        \captionsetup{justification=raggedright, singlelinecheck=false, position=top, aboveskip=-1pt, belowskip=-1pt}
        \caption*{(b)}
        \includegraphics[width=\textwidth, height=0.9\textwidth]{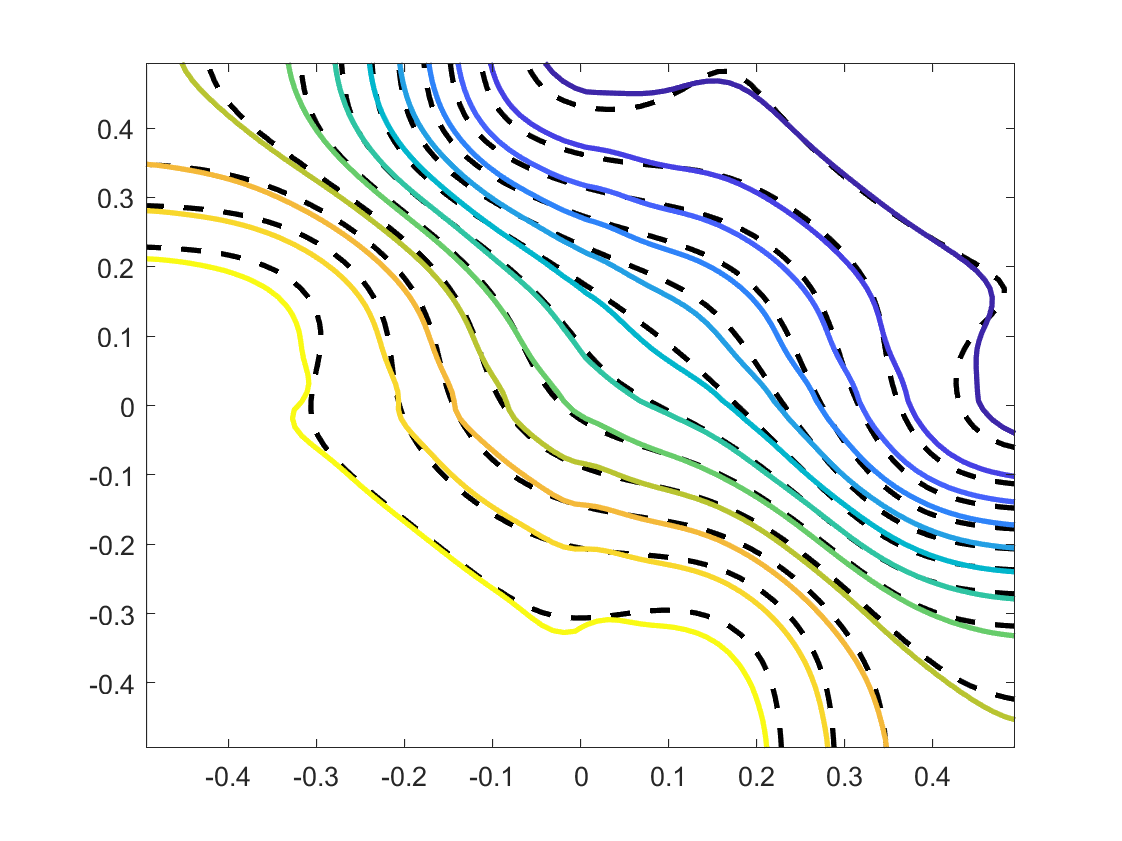} 
    \end{subfigure}
    \hfill
    \begin{subfigure}[b]{0.32\textwidth}
        \centering
        \captionsetup{justification=raggedright, singlelinecheck=false, position=top, aboveskip=-1pt, belowskip=-1pt}
        \caption*{(c)}
        \includegraphics[width=\textwidth, height=0.9\textwidth]{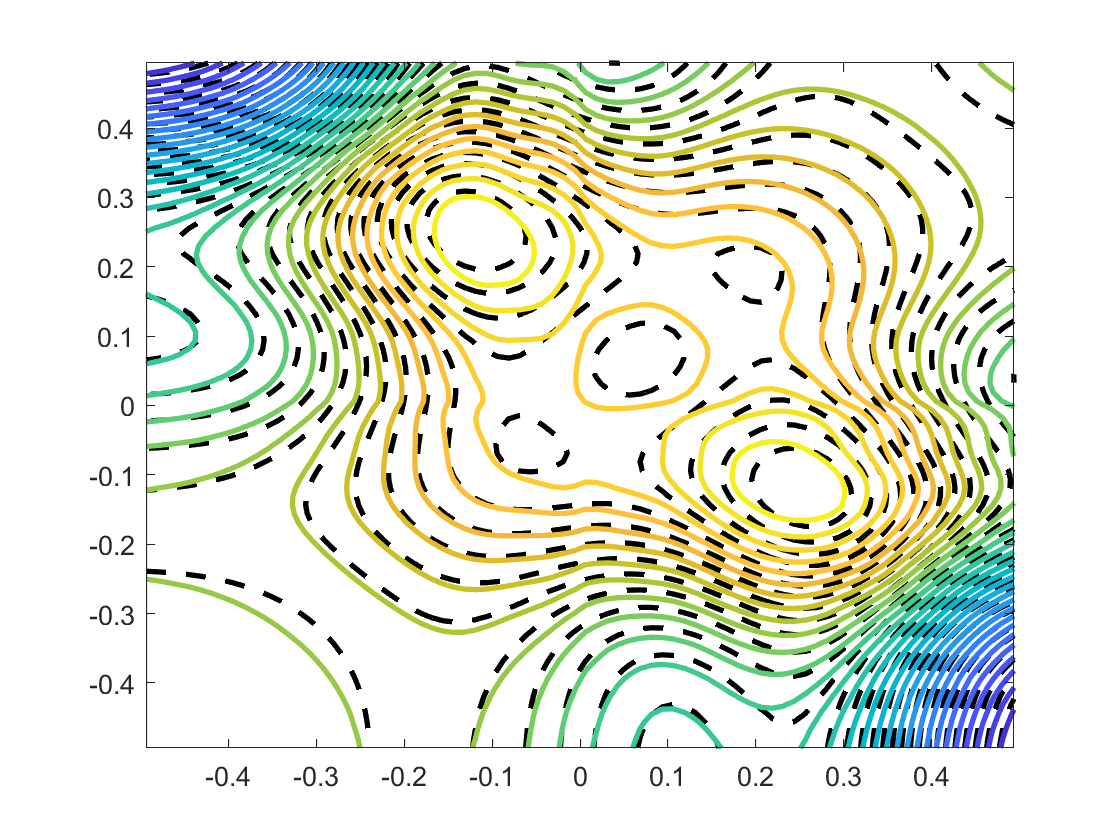} 
    \end{subfigure}
    \caption{2-D Riemann problem with $\tau = 10^4$: contours of (a) density $\rho$, (b) velocity magnitude $\bm{U}$ and (c) temperature $\theta$ at $t = 0.15$. The black dashed lines, analytical solutions. The colorful lines, numerical solutions given by the DVD-HyQMOM model. We set $n = 8$ and $N = 10$, with directions $\left\{\bm{l}_m = \left(\cos\frac{(2m-1)\pi}{20}, \sin\frac{(2m-1)\pi}{20}\right)^{T}\right\}^{10}_{m = 1}$.}
    \label{2dcoless_4}
\end{figure}

For the free-molecular regime, we set $n = 8$ and $N = 10$ for DVD-HyQMOM, and the directions are $\left\{\bm{l}_m = \left(\cos\frac{(2m-1)\pi}{20}, \sin\frac{(2m-1)\pi}{20}\right)^{T}\right\}^{10}_{m = 1}$. Figure \ref{2dcoless_4} exhibits the contours of density $\rho$, velocity magnitude $\bm{U}$, and temperature $\theta$ at $t = 0.15$, together with the analytical solutions \cite{xu15}. It can be observed that increasing the number $n$ significantly improves the accuracy of the results compared with small $n$ cases and other moment methods, especially for the density and velocity magnitude. However, a limitation remains in the precision of the temperature profiles, implying that further refinement of higher-order terms and improvements of our numerical schemes may still need to be considered.

\begin{figure}[h]
    \centering
    \includegraphics[width=0.6\textwidth, height=0.45\textwidth]{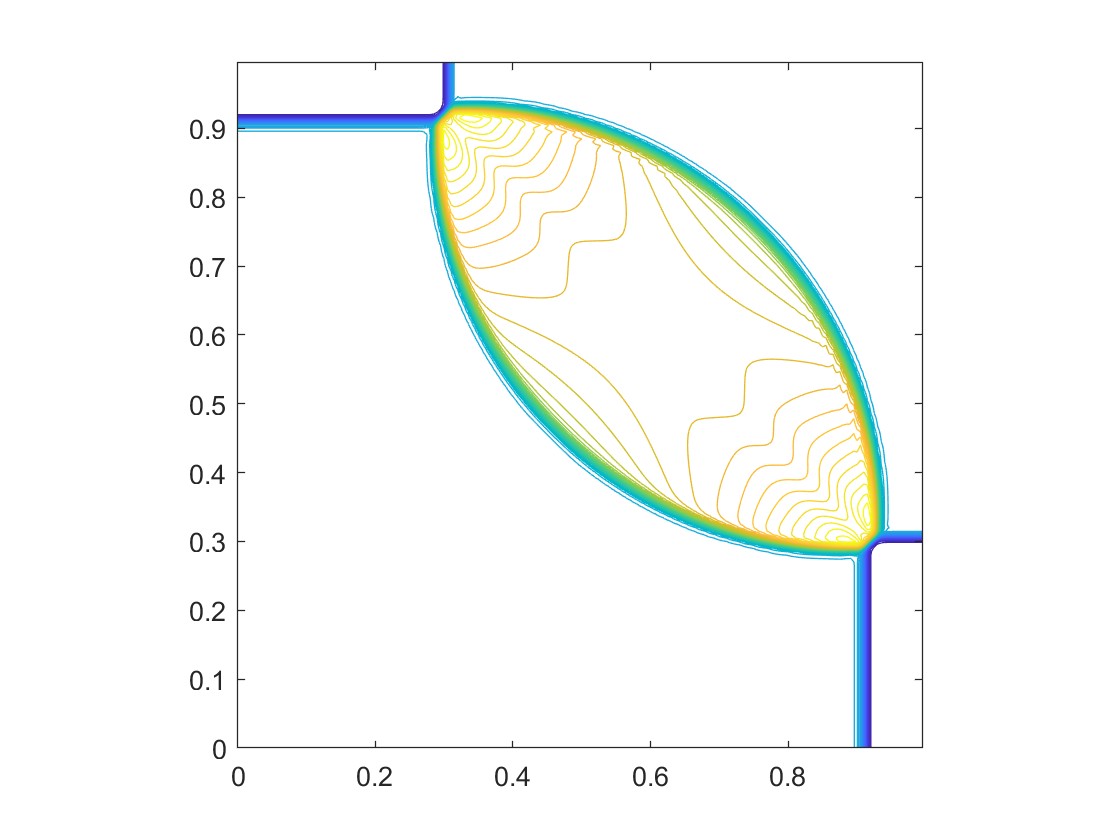}
    \caption{2-D Riemann problem with $\tau = 10^{-4}$: density contour $\rho$ at $t = 0.25$, utilizing the DVD-HyQMOM model. We set $n = 2$ and $N = 8$, with directions $\left\{\bm{l}_m = \left(\cos\frac{(2m-1)\pi}{16}, \sin\frac{(2m-1)\pi}{16}\right)^{T}\right\}^{8}_{m = 1}$.} 
    \label{2dcol_4}
\end{figure}

For the continuum regime, we set $n = 2$ and $N = 8$ for DVD-HyQMOM, and the correlated directions are $\left\{\bm{l}_m = \left(\cos\frac{(2m-1)\pi}{16}, \sin\frac{(2m-1)\pi}{16}\right)^{T}\right\}^{8}_{m = 1}$. Figure \ref{2dcol_4}
shows the density contour at time $t = 0.25$, which agrees quite well with the solutions shown in \cite{liu98}. It is worth noting that, in this case, we employed the third-order energy stable WENO (ES-WENO) scheme to better achieve the DVD-HyQMOM model's effectiveness, rather than using the first-order 2-D upwind scheme mentioned in the previous section.

In order to check the robustness of the DVD-HyQMOM model, we apply it to the 2-D Riemann problem with another initial conditions \cite{xu15, liu98}:
\begin{equation*}
    (\rho, u, v, p) = \left\{ 
  \begin{array}{lll}
    (\rho_1, u_1, v_1, p_1) = (0.5313, 0, 0, 0.4)& x > 0, & y>0\\
     (\rho_2, u_2, v_2, p_2) = (1, 0.7276, 0, 1)& x\le 0, & y>0\\
     (\rho_3, u_3, v_3, p_3) = (0.8, 0, 0, 1)& x\le 0, & y\le 0\\
     (\rho_4, u_4, v_4, p_4) = (1, 0, 0.7276, 1)& x> 0, & y\le 0\\
  \end{array}
\right.
\end{equation*}

\begin{figure}[ht]
    \centering
    \begin{subfigure}[b]{0.32\textwidth}
        \centering
        \captionsetup{justification=raggedright, singlelinecheck=false, position=top, aboveskip=-1pt, belowskip=-1pt}
        \caption{}
        \includegraphics[width=\textwidth, height=0.9\textwidth]{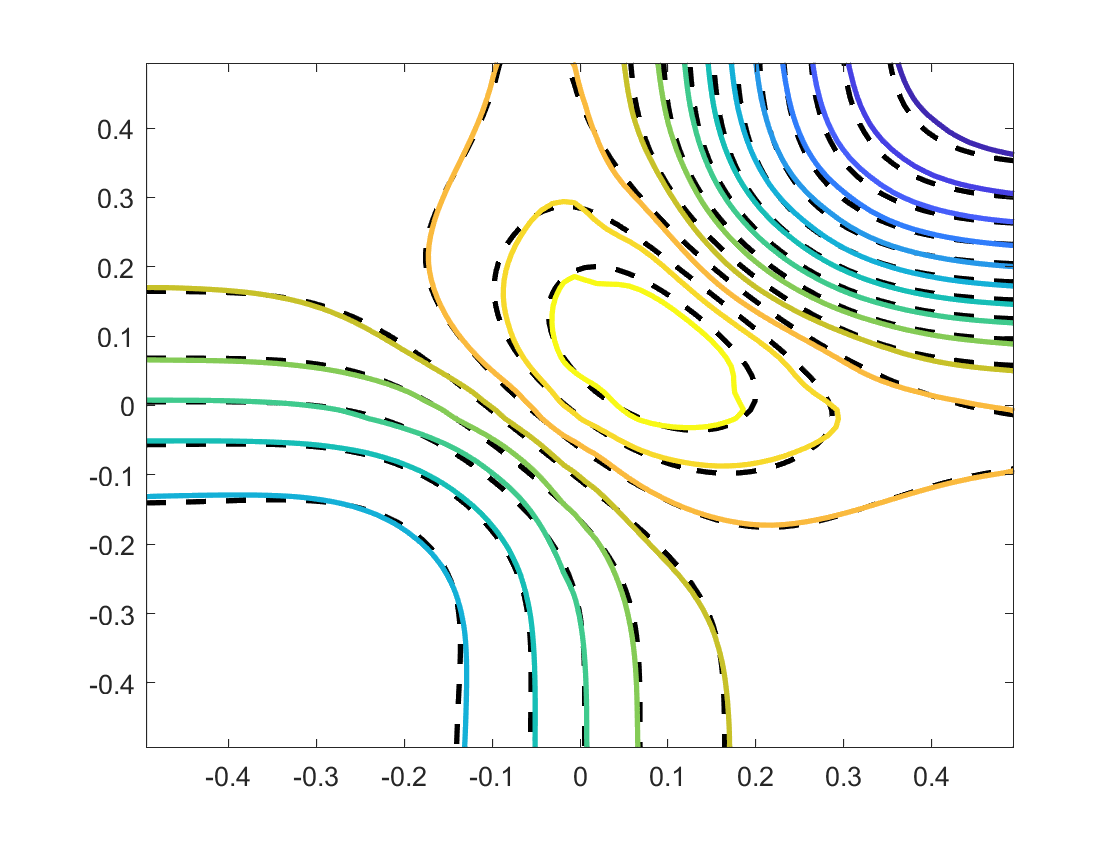}
    \end{subfigure}
    \hfill
    \begin{subfigure}[b]{0.32\textwidth}
        \centering
        \captionsetup{justification=raggedright, singlelinecheck=false, position=top, aboveskip=-1pt, belowskip=-1pt}
        \caption{}
        \includegraphics[width=\textwidth, height=0.9\textwidth]{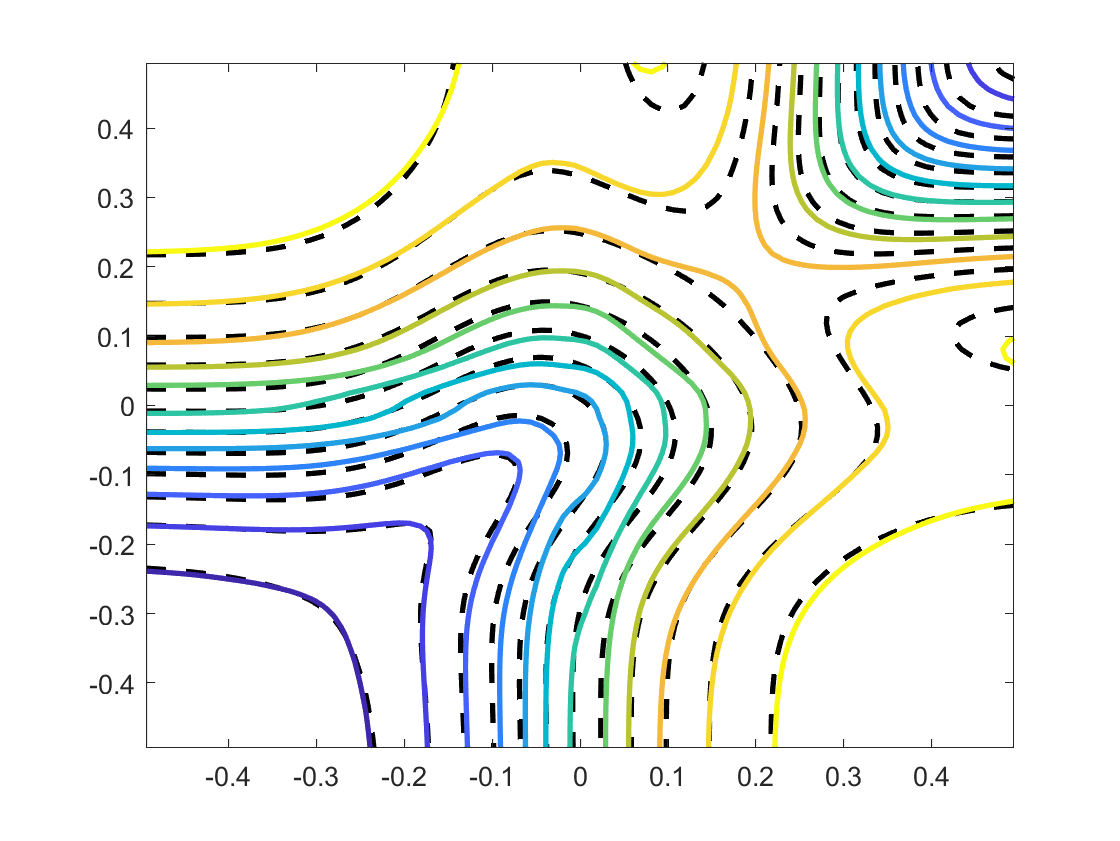} 
    \end{subfigure}
    \hfill
    \begin{subfigure}[b]{0.32\textwidth}
        \centering
        \captionsetup{justification=raggedright, singlelinecheck=false, position=top, aboveskip=-1pt, belowskip=-1pt}
        \caption{}
        \includegraphics[width=\textwidth, height=0.9\textwidth]{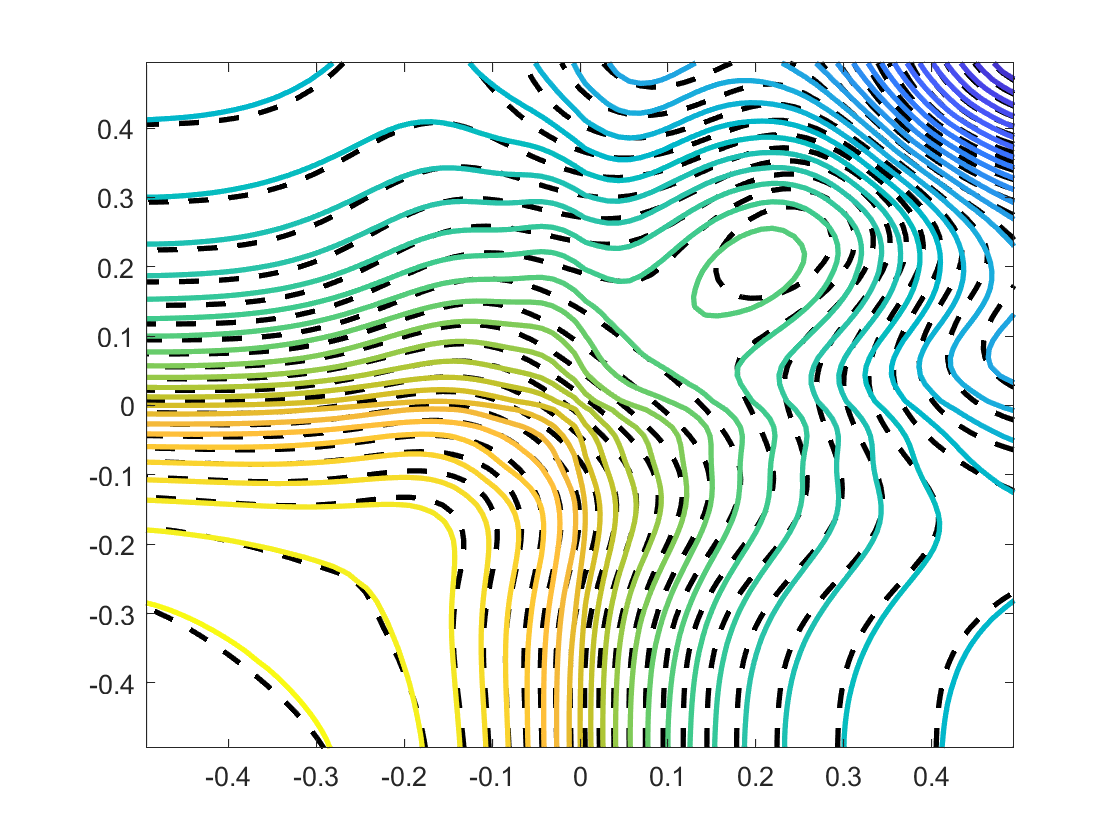} 
    \end{subfigure}
    \caption{2-D Riemann problem with $\tau = 10^4$: contours of (a) density $\rho$, (b) velocity magnitude $\bm{U}$ and (c) temperature $\theta$ at $t = 0.15$. The black dashed lines, analytical solutions. The colorful lines, numerical solutions given by the DVD-HyQMOM model. We set $n = 8$ and $N = 10$, with directions similar as before.}
    \label{2dcoless_12}
\end{figure}

We only check the free-molecular regime, and the results of density, velocity magnitude, and temperature are shown in figure \ref{2dcoless_12}. Here we set $n = 8$ and $N = 10$ again, with the same uniform cells and directions as the former case.

\section{Conclusions} \label{sec: con}
In this paper, we have applied the discrete-velocity-direction model (DVDM) on the hyperbolic quadrature method of moments (HyQMOM), developing a promising multidimensional spatial-time approximation of the original BGK equation, DVD-HyQMOM. It can be regarded as the multidimensional version of HyQMOM, presenting higher accuracy compared with other DVDM submodels, espscially when the number of abscissas is large enough. 

The feasibility of DVD-HyQMOM has been verified numerically, using the first-order upwind scheme. Here the Neumann boundary condition and the solid boundary condition with diffuse-scattering law have been applied on Riemann problems and Couette flow problems respectively. 
The numerical results for 1-D and 2-D Riemann problems, particularly in both hydrodynamic and rarefied regimes, demonstrate the capability of the DVDM submodels to capture flow discontinuities effectively.
Additionally, the results for Couette flow problems show strong agreement with benchmark data across a broad spectrum of flow regimes.

This model currently has certain limitations, particularly in the weight-solving processes. Therefore, identifying more effective methods to increase the number of abscissas, while also improving algorithms to reduce runtime as the abscissas count grows, presents a promising direction for future work.


\nocite{*} 

\end{document}